\documentclass[journal]{IEEEtran}

\ifCLASSINFOpdf

\else

\fi

\usepackage{cite}
\usepackage{amsmath,amssymb,amsfonts}
\usepackage{algorithmic}
\usepackage{graphicx}
\usepackage{textcomp}
\usepackage{verbatim}
\usepackage[ruled, vlined]{algorithm2e}
\usepackage{mathtools, cuted}
\usepackage{hyperref}
\hypersetup{
    colorlinks=true,
    linkcolor=black,
    filecolor=black,     
    citecolor=black,
    urlcolor=blue,
}

\def\BibTeX{{\rm B\kern-.05em{\sc i\kern-.025em b}\kern-.08em
    T\kern-.1667em\lower.7ex\hbox{E}\kern-.125emX}}

\begin{document}

\title{Blockchain-Enabled IoT Platform for 
End-to-End Supply Chain Management}

\author{Keqi~Wang,
        Wei~Xie,
        Wencen~Wu,
        Jinxiang~Pei,
        and~Qi~Zhou
\thanks{Corresponding author: Wei Xie (e-mail: w.xie@northeastern.edu).}
\thanks{Keqi Wang, Wei Xie and Jinxiang Pei are with Mechanical and Industrial Engineering, Northeastern University
Boston, MA 02115, USA (e-mail: wang.keq@northeastern.edu; w.xie@northeastern.edu; ji.pei@northeastern.edu).}
\thanks{Wencen Wu is Computer Engineering Department, San Jose State University,
San Jose, CA 95192, USA (e-mail: wencen.wu@sjsu.edu).}
\thanks{Qi~Zhou is QuarkChain Inc., San Mateo, CA 94401, USA (e-mail: qizhou@quarkchain.us).}
}

\maketitle

\vspace{-0.2in}

\begin{abstract}
In this paper, we propose a blockchain-enabled IoT platform to support interoperability and traceability, improve the efficiency and throughput of quality control verification, and ensure the safety of end-to-end supply chain. Even though it is motivated by industrial hemp industry, the platform can be applicable to general biopharmaceuticals, agriculture and food supply chains.
After being legalized as an agricultural commodity by the 2018 U.S. Farm Bill, the Industrial Hemp production is moved from limited pilot programs to a regulated agriculture production system, and the market keeps increasing since then. However, Industrial Hemp Supply Chain (IHSC) faces several critical challenges, including high complexity and variability, data tampering, and lack of an immutable information tracking system. In this paper, we develop a blockchain-enabled internet-of-things (IoT) platform for IHSC to support process tracking, scalability, interoperability, and risk management. Basically, we utilize the parallel processing and create a two-layer blockchain with proof-of-authority based smart contracts, which can leverage local authorities with state/federal regulators to ensure and accelerate quality control verification and regulatory compliance. 
Then, we develop a user-friendly mobile app so that each participant can use a smart phone to real-time collect and upload their data to the cloud, and further share the process veriﬁcation and tracking information through the blockchain network. Thus, the proposed blockchain-enabled IoT platform can support interoperability and traceability, improve the efficiency and throughput of quality control verification, and ensure the safety of regulated IHSC.
\end{abstract}


\section*{Managerial relevance statement}

We develop a blockchain-enabled internet-of-things (IoT) platform for the Industrial Hemp Supply Chain (IHSC) to support process tracking, scalability, interoperability, and improve safety and efficiency. It can tackle the critical challenges of IHSC quality control verification and risk management, including high complexity and variability, data tampering, lack of an immutable information tracking system, and increasing market. Basically, we develop a two-layer blockchain with the proof-of-authority based smart contracts and a hierarchical verification procedure. It can leverage the resource from local authorities and allow multiple shard chains simultaneously and promptly process the transactions coming from different areas. The proposed validation mechanism can ensure product quality control and regulatory compliance.
In addition, we develop a user-friendly mobile app, which can facilitate flexible, real-time information collection, verification and sharing, and tracking. The proposed platform can be easily extended to a wide set of highly regulated supply chain management, such as global biopharmaceutical  manufacturing and supply chain.

%
\IEEEpeerreviewmaketitle

\section{Introduction}
\label{sec:introduction}

\textit{The {objective} of our study is to create a blockchain-enabled IoT platform for integrated industrial hemp supply chain (IHSC), which can improve the traceability, accelerate the regulation verification and quality control, and facilitate the development of a safe, efficient, reliable, and automated supply chain system.} 
Even though the proposed platform is general, in this paper, we consider the end-to-end IHSC, including seed breeding, cultivation, pre-harvest test and harvest, stabilization, processing, refinement of products/by-products, and storage/transportation to the end users.

After being legalized as an agricultural commodity by the 2018 U.S. Farm Bill, industrial hemp (IH) production is moved from limited pilot programs administered by state regulatory officials and becomes a public \textit{regulated} agriculture production. Due to the huge success of farmers who participate in the pilot programs in various states, there is intense interest in growing IH, and the total area licensed for hemp production has increased dramatically from 37,122 (in year 2017) to 310,721 (in year 2019) acres nationally 
\cite{sterns2019emerging}. According to the U.S. Department of Agriculture (USDA), the sales are expected to increase from \$25 million in 2020 to more than \$100 million by 2022. 

 The two most common cannabinoid produced from hemp are cannabidiol (CBD mainly for medicinal uses) and tetrahydrocannabinol (THC mainly for recreational drug use). As opposed to the safety of CBD, THC has been reported some unpleasant side-effects including but not limited to anxiety and panic, impaired attention, memory, and psychomotor performance while intoxicated \cite{hall1998adverse}. Considering the major active ingredient in marijuana causing psychoactive effects, 
most state regulatory officials set the permissible THC concentration levels as not exceed 0.3 percent on a dry weight basis and laboratories must test for and report it; see \href{https://oregon.public.law/rules/oar_chapter_603_division_48}{Division 48 Industrial Hemp}. 
\textit{Thus, the development of a safe, reliable, sustainable, efficient and automated IHSC plays a critical role in improving the economy and ensuring public health.}

However, the management of IH industry faces critical challenges, including 
scalability, high complexity and variability, very limited IHSC knowledge, data tampering, and lack of immutable data/information tracking system. First, the production process of end-to-end IHSC is complex and 
many factors can contribute to the CBD and THC production and dynamic flows.
Second, there isn't a well-developed platform to track historical records and maintain the immutability and transparency of data. Third, the veracity of the collected data is hard to guarantee, especially for the data required by the regulation. Fourth, how to allocate the limited inspection resources from state and federal regulatory officials to monitor this fast-growing industry is not clear.

\textit{Therefore, IH practitioners and government regulators have urgent needs, including improving traceability and transparency, 
eliminating the risks, ensuring product quality, and controlling the regulatory THC/CBD production and flow through the supply chain.}
A potential solution to alleviate all these challenges and concerns in the blockchain technology, which is a peer-to-peer digital ledger built on blockchain network rather than relying on centralized servers. 
Even though the conventional one-layer blockchain can support traceability, immutability, transparency, and regulatory compliance, the process validation efficiency is limited by the single chain capacity and the number of available state regulation officials for on-site visit and quality verification. 
Given the restricted capacity of a single blockchain and the limited state/federal regulator inspection resources at each state, 
it is challenging to handle the situation when the scale of IH industry grows dramatically.

Here, we summarize the key contributions of this paper. Driven by the critical needs from IHSC quality control and risk management, we first create a two-layer blockchain platform and design a proof-of-authority-based smart contract, which explores the parallel computing and utilizes the state sharding technique to simultaneously process the transactions from different areas. It can leverage the additional resources from local authorities 
(e.g., local quality assurance consulting and service companies) with state/federal regulators to improve the efficiency and safety of IHSC.
In addition, we develop a user-friendly mobile app so that each participant can use a smart phone to real-time collect and upload their validated data to the cloud, share the process veriﬁcation results, and support the historical information/record online tracking through the blockchain network, which can improve the flexibility and efficiency of IHSC.

This study is a foundation of our academia-industry collaboration on {``AI- and Blockchain-based IoT Platform Development for End-to-End IHSC Learning, Risk Management, and Automation."}  Our blockchain is built on QuackChain network (see \href{https://github.com/QuarkChain/pyquarkchain/wiki/Cluster-Design}{quarkchain wiki}). 
{By collaborating with multiple research teams from Oregon State University, Alabama A\&M University, Colorado State University-Pueblo, Colorado State University, and industrial partners (i.e., Willamette Valley Assured LLC), we test and validate our developed blockchain-enabled IoT platform during the real-world small-scale pilot phase: IH season 2020 
in different states.} 

This paper is organized as follows. In Section~\ref{sec:review}, we review the most related blockchain development and applications. In Section~\ref{sec:IHSC}, we describe the end-to-end industrial hemp supply chain (IHSC), main participants, and their expectation and needs. In Section~\ref{sec:blockchain_IoT}, we propose the blockchain-enabled internet-of-things (IoT) platform, including IHSC process monitoring and tracking, two-layer blockchain design, 
smart contract, and consensus design for data/information sharing, process quality validation, THC/CBD control, and regulation compliance. In Section~\ref{sec:implementation}, we discuss the architecture of the proposed platform, and present the key algorithms for record creating, verification, validation, and data/information retrieving. Then, the performance of the proposed blockchain-enabled IoT platform for improving the IHSC efficiency and safety is studied, and the real-world implementation at Oregon, Alabama, Colorado, and Pennsylvania 
is discussed in Section~\ref{sec:performance}. We conclude this paper in Section~\ref{sec:conclusion}.

\section{Literature Review}
\label{sec:review}

Here, we briefly summarize the most relative literature. For more comprehensive literature review on blockchain development and application, please refer to \cite{8731639, kamilaris2019rise, alladi2019blockchain}.
Blockchain technology was first introduced by Satoshi Nakamoto as Bitcoin in 2008 to solve the double-spending problem \cite{nakamoto2019bitcoin}. Beyond Bitcoin, the so-called second generation of cryptocurrencies, Ethereum, was proposed in \cite{wood2014ethereum, buterin2014next}. Ethereum is a blockchain with a built-in Turing-complete programming language, which allows the developers to write smart contracts. 
A number of decentralized blockchain applications have already been built based on it; see for example \cite{alladi2019blockchain,soltanisehat2020technical}. 

Blockchain is a distributed ledger technology that 
maintains all transactions between two or more involved participants through a peer to peer (P2P) network. 
As presented in \cite{8386948}, it can work as a permanently immutable ledger 
upon blockchain network, no matter the involved participants (nodes) are trustful or not. The stored records are co-owned by all members of the network.
Without a central controller in the blockchain architecture, before a new block is appended to the existing blockchain, all the nodes have to reach a consensus. 
Many different consensus algorithms have been proposed, including Proof-of-Work (PoW) \cite{nakamoto2019bitcoin}, Proof-of-Stake (PoS) \cite{king2012ppcoin}, Proof-of-Authority (PoA) \cite{de2018pbft},  Proof-of-Object (PoO) \cite{mondal2019blockchain}, Ripple Protocol Consensus Algorithm (RPCA) \cite{schwartz2014ripple}, etc.  


Although the main focus of classical blockchain applications includes banking, finance, and insurance industries, the attempts to explore and extend to various other domains (e.g., supply chain, healthcare, and agriculture) and start to gain popularity.
One of the most crucial challenges of supply chain risk management is tracking the data provenance and maintaining its traceability and transparency throughout the integrated and end-to-end supply chain network. The traditional supply chains are centralized, which depends on a third party to validate the trading. This proves vulnerable to both data modification and management, and also it impacts the process efficiency and interoperability. The literature \cite{sander2018acceptance} 
shows that the blockchain technologies can improve food safety by enhancing transparency. 
In \cite{bettin2018methodological}, the authors integrate the blockchain technology into the food supply chain management, which allows the traceability along the process and provides the end customer with enough information about the origin of product in order to make an informed purchase decision. 
Through the blockchain, the grain quality assurance is created and shared transparently to the public, which can lead to an added valuation of its selling price. 
The authors in \cite{8718621} utilize Ethereum blockchain and smart contracts to control all interactions and transactions among the participants involved in the soybean supply chain ecosystem without the validation from a trusted third-party authority. 
The study in \cite{9058674} presents  a complete blockchain-based solution to ensure traceability, trust, and delivery mechanism in Agriculture and Food (Agri-Food) supply chain. 
Multiple smart contracts 
are designed to support the interactions of different types of entities 
in the system. 
The authors in \cite{osmanoglu2020effective} propose a blockchain-based solution that carries out the yield estimation of agricultural products. Thus, the necessary precautions for the excessive imbalances that may arise in agricultural products will be planned in advance and the risk of price fluctuation would be hedged. In  \cite{du2020supply}, the authors develop a new blockchain-enabled supply chain financial platform, which can 
improve the efficiency of the capital and information flow, reduce the cost, and provide better financial services to the relevant parties in the supply chain. 

Due to its innovative features, blockchain technology can be used to facilitate Internet of Things (IoT) and Industrial IoT (IIoT) \cite{wang2020blockchain}. Reference \cite{7987376} uses IoT sensor devices leveraging blockchain technology and Ethereum smart contracts to assert data immutability and public accessibility of environmental records (e.g., temperature and humidity) over the transport of medical products. The sensor devices can monitor the storage environment during the shipment to ensure the regulations compliance on ``Good Distribution Practice of medicinal products for human use (GDP)".
Another work \cite{mondal2019blockchain} creates an end-to-end blockchain architecture for food traceability by integrating the radio frequency identification (RFID)-based sensors at the physical layer and  blockchain at the cyber layer. The authors implement a proof-of-object-based authentication protocol to ensure network security and reduce the cost. Researchers in \cite{caro2018blockchain} present AgriBlockIoT, a fully decentralized, blockchain-based traceability solution for Agri-Food supply chain management, which is able to integrate various IoT devices and directly produce the valuable information along the whole supply chain. The implementations of AgriBlockIoT are illustrated on both Ethereum and Hyperledger platforms. Meanwhile, 
built on the blockchain and Electronic Product Code Information Services (EPCIS) network, \cite{lin2019food} proposes a decentralized system 
for food safety traceability. A prototype of the proposed architecture has been implemented to demonstrate the effectiveness and superiority over the existing systems. The system consists of on- and off-chain modules, which can alleviate the data explosion issue of the blockchain for the IoT.

As far as we know, there is a lack of  blockchain literature to improve the traceability, quality, and safety of regulated end-to-end industrial hemp supply chain. Therefore, in this paper, an IHSC blockchain-enabled IoT platform is developed. The performance analysis and implementation of this platform are also demonstrated. 

\section{End-to-End Industrial Hemp Supply Chain}
\label{sec:IHSC}

Even though the proposed blockchain-enabled IoT platform can be extended to general agri-food and biopharmaceutical supply chains, in this paper, we focus on the IHSC from seed selection to final commercial product, called ``THC free broad-spectrum CBD oil", which will be shortened as \textit{CBD oil} for simplification. In this section, we introduce the end-to-end industrial hemp supply chain.
We describe the CBD oil based IHSC procedure in Section~\ref{subsec:IHSC_procedure}, summarize the main groups or categories of participants involved in the process, and discuss their expectations/needs in Section~\ref{subsec:participants}.

\begin{figure*}[htb]
{
\centering
\includegraphics[width=\textwidth]{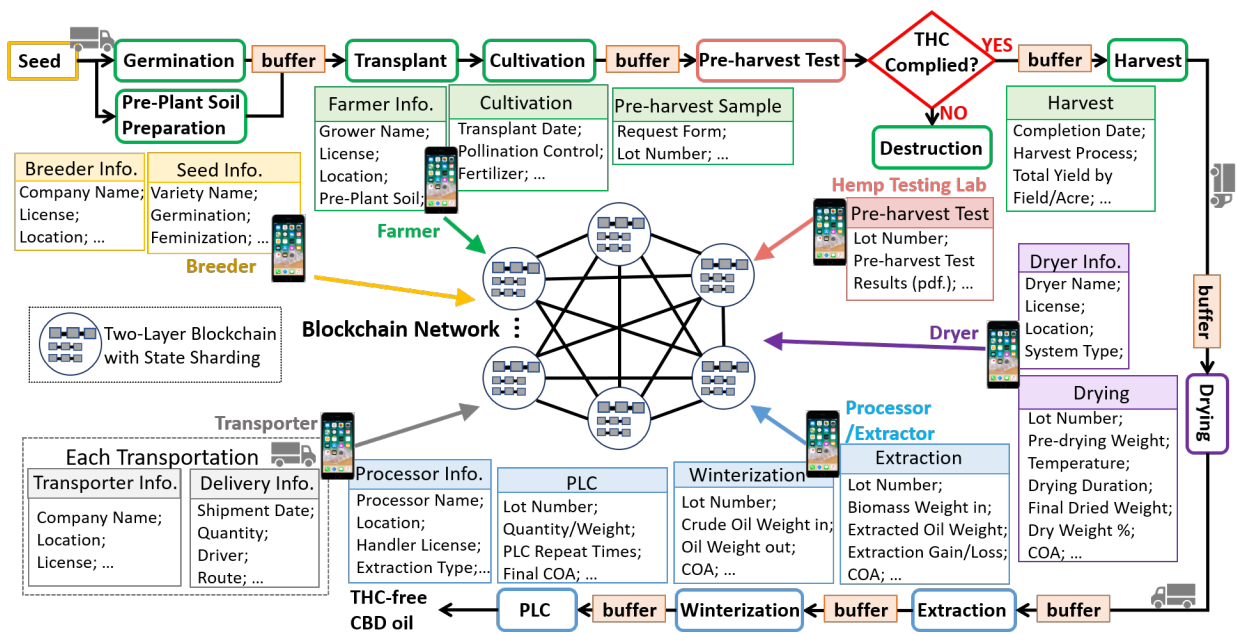}
\vspace{-0.6in}
\caption{Illustration of Blockchain-Enable End-to-End Industrial Hemp Supply Chain \label{fig: IHSC}}
}
\end{figure*}

\subsection{IHSC Process Introduction}
\label{subsec:IHSC_procedure}

The \textit{main operating units of IHSC include: (1) cultivation, (2) stabilization (drying), and (3) manufacturing (extraction, winterization, and purification-PLC); see Figure~\ref{fig: IHSC}.} 

\noindent \underline{\textbf{Cultivation:}} 
It starts with \textit{seed sourcing}, including the selection of seed variety, and licensed growers (farmers) select seed varieties to grow for the incoming season. Once receiving the selected seed, the farmer simultaneously starts Germination and Pre-Plant Soil Preparation process. \textit{Germination} happens at the greenhouse with the controlled-growth environment, where the seed germinates (the leaves shoot up and the roots dig down) in about 5--10 days. After germination, there is roughly a 2-day holding period restriction before transplanting. 
It is important that the seedlings are not stressed after leaving the greenhouse and being planted. In the meantime, after the \textit{Soil Test} (which provides a detailed and comprehensive description of soil, such as pH, nutrition level, heavy metal, herbicide, and pesticide contains), the farmer performs \textit{Pre-Plant Soil Preparation} to adjust the properties of field soil, and this process usually takes 1--2 days. Once both steps are finished, the germinated seedlings are transplanted into the field, which takes about 1--2 days. 
Then, the farmer continues the cultivation process including two typical sub-phases: Vegetative and Flowering. \textit{Vegetative} is the main growth stage, where the iconic fan leaves begin to develop and grow larger with more blades and fuller sets of leaves. \textit{Flowering} is the final stage of growth, where the resinous buds develop (the majority of the CBD is generated). Once the flowers begin to develop, they require a different kind of care. Besides the regular control of weed, insect, and mold, the most important one is pollination control. If the plants are pollinated, they shift their energy from producing unpollinated flowers that are high in oil to pollinated flowers that consume oil to produce seed, which will decrease the CBD yield significantly. Overall, the cultivation process in total takes about 50--60 days. 

After that, the sample of IH is sent to Hemp Testing Lab for \textit{pre-harvest test} that takes 2--7 days on average and returns the comprehensive test results including CBD/THC levels. \textit{Based on the state regulation, the lot with THC level greater than 0.3\% is required to be destroyed completely.} For the IH lots that have passed the pre-harvest test, farmers can start the \textit{harvest process}. Usually, farmers tend to wait for several days since the CBD level further increases during this period. However, since the THC content also increases, the government requires farmers to complete harvest no longer than 15 days after the pre-harvest test. 
If farmers don't finish harvesting these lots in time, the rest needs to be scheduled for another testing and then followed by new harvest within another 15-day limit. Thus, farmers need to make an appropriate harvest schedule to consider the risk-profit trade-off.

\noindent \underline{\textbf{Stabilization:}} 
The harvested IH biomass often has a high moisture content, even up to 85\%, which has to be reduced to an ideal level to avoid deterioration in 
storage and transportation. Typically, there are two types of stabilization: \textit{drying-stabilizing} and freezing-stabilizing. Since drying is most common for the post-harvest stabilization process, we consider it here. Based on the drying machine availability, the harvested IH biomass may have to wait. The probability of being contaminated increases during waiting with high moisture, which can cause loss in profit. The drying process typically takes 1-2 days.

\noindent \underline{\textbf{Manufacturing:}}
The dried IH biomass is then sold to licensed processors/extractors to produce the desired commercial product (CBD oil). The main steps include Extraction, Winterization, and Preparatory Liquid Chromatography (PLC). Since the dried biomass has a life of 9--12 months, the product can wait in a buffer before each step. \textit{Extraction} is a process to produce crude oil, which typically includes two steps: extraction and decarboxylation. 
The Ethanol Extraction is one of the most efficient  methods for processing large batches of IH biomass. This method cycles ethanol solvent through the solid biomass, strips the cannabinoids, terpenes, and plant waxes from the flower to produce the CBD crude oil. Although CBD is the main cannabinoids with the healthy and medicinal potential,  
the highest concentrations of cannabinoid in biomass (flowers) is cannabidiolic aid (CBDA). After the extraction process, the decarboxylation reaction is needed to transform CBDA into its neutral cannabinoids CBD  \cite{wang2016decarboxylation}. Also, the decarboxylation can prevent the degradation of desirable cannabinoids \cite{romano2013cannabis}. Thus, to decarboxylate the CBDA to CBD, the oil is to be processed under controlled temperature. In the meantime, the tetrahydrocannabinolic acid (THCA) is transformed as THC. Not only the cannabinoids but also other undesirable elements in biomass are extracted. The whole Extraction process usually takes 1--2 days. Then, another process, called \textit{winterization}, needs to be applied to remove the undesirable fat, terpene, and wax from crude oil. During this process, the content of the CBD/THC stays almost the same with very minor drop in absolute weight, and the percentage of CBD/THC increases. This process takes about 1--3 days. Finally, several \textit{PLC steps} are used to remove some specific target molecule, such as THC, and remediate the oil with THC level below the standardized requirement. 
After the PLC steps, there will be another Certificate of Analysis (COA) test, and the final product can only be acceptable with less than 0.05\% THC (for THC free broad-spectrum CBD oil).

\subsection{Description of Regulated Participants}
\label{subsec:participants}

The practitioners and government regulators have urgent needs, including: improving the transparency, developing a comprehensive and deep understanding of end-to-end IHSC, reducing the risks, ensuring the CBD products quality, and controlling the THC production and flow through the supply chain process. More detailed expectation from different {participants} is listed as follows.
\begin{enumerate}
    \item \underline{\textbf{Breeder:}} A breeder produces IH seed varieties or clones, which will be sold to the farmers. They want to obtain the real data about their varieties' attributes and performance (e.g., yield, mold resistance) outside of lab across different locations under various
    environmental conditions, which can facilitate determining the optimal adaptation of hemp essential oil variety types and genetics across U.S. farm resource regions. 
    \item \underline{\textbf{Licensed Grower:}} A grower obtains the seeds varieties from breeder, and then cultivates the IH until harvest. The performance of seed variety (e.g., yield, THC/CBD content, etc.) is the key criterion when they purchase the seed. During the cultivation, they need professional suggestions to cultivate their crop. After harvest, they need to find reliable and stable buyers. In addition, they are interested in keeping tracking their IH flow to the end of the supply chain.
    \item \underline{\textbf{Dryer:}} A dryer provides the harvested IH drying-stabilization service for the growers. They can properly schedule the equipment based on the updating volume information about harvested IH.
    \item \underline{\textbf{Licensed Processor/Extractor:}} A processor purchases the dried IH biomass from the growers/dryers, and then conducts processing and extraction to produce the commercial CBD oil. He/she wants to find high quality and stable IH supplier. 
    \item \underline{\textbf{Transporter:}} A transporter ships and delivers the items  (intermediate IH product) between different participants of the IHSC. Due to the legality of THC content control, sharing the credible information IH shipment with regulators and authorities will help the transporter pass the checking points faster.
    \item \underline{\textbf{Hemp Testing Lab:}} A Hemp Testing Lab receives the IH samples from the official sampler and then provides the pre-harvest test results to the growers.
    \item \underline{\textbf{Authorities/Regulators:}} The local authorities are responsible for the verification in their local area, and then the regulators (i.e., state USDA) will confirm all the uploaded data (may randomly select some for detailed verification). The majority of their duty is monitoring and surveilling the whole supply chain, tracking the THC, and solving the dispute.
\end{enumerate}

\noindent To meet the unique expectations and needs from different participants, the traceability, accountability, credibility, and auditability are the key requirements for IHSC management.



\section{Blockchain-enabled IoT Platform}
\label{sec:blockchain_IoT}

Here, we present the proposed blockchain-enabled IoT platform for the end-to-end IHSC to improve the transparency, safety, efficiency, and throughput. In Section~\ref{subsec:processMonitoring}, we describe IHSC process monitoring and CBD/THC tracking. We discuss the  critical process parameters/information and regulation required data, which can support the process tracking, learning, interoperability, decision making, and real-time problem detection.
Then, built on QuarkChain introduced by Qi Zhou (see \href{https://github.com/QuarkChain/pyquarkchain/wiki/Cluster-Design}{quarkchain wiki}),
for the highly \textit{regulated IHSC}, we present the proposed two-layer blockchain, proof-of-authority (PoA) based smart contract, and hierarchical consensus design in Sections~\ref{subsec:blockchain} and \ref{subsec:consensus}. 

This study allows us to leverage the additional verification resources from local authorities to assist the regulatory agency (i.e., USDA), which can improve the IHSC safety, quality control, efficiency, and scalability. We design the geography-based state sharding and two-layer blockchain, which can simultaneously process the jobs coming from different areas. 
In addition, to improve the transparency and safety and facilitate the product quality verification for the end-to-end IHSC, we develop a user-friendly mobile app so that the internet-connected smartphones or devices can be used to real-time dynamically monitor the process, record and share the important information, and track the historical records/data.

\subsection{IHSC Process Monitoring and Tracking}
\label{subsec:processMonitoring}


For the highly regulated industrial hemp industry, all participants require to be {licensed}. When they upload the information to the blockchain-enabled IoT system and database, their licenses and background need to be verified online. 
\textit{We summarize the critical information of participants in Table~\ref{tab:IEEEPartInfo}, including name, background, and permission in terms of their activities in the IHSC ecosystem.}

\textit{The transactions occurring in each operating unit of the end-to-end IHSC are categorized and described in Table~\ref{tab:IEEEProInfo}.} 
For each transaction $\mathbb{T}$, the signature $\mathbb{S}$ records the participant(s) who will take the responsibility for the accuracy and truth of the information uploaded to the blockchain and database system.
The information $\mathbb{I}$ column summarizes the critical information and data needed to be recorded, including the messages required by the regulations and other important data used to track the IHSC process.

For all the uploaded information and data, we have online verification $\mathbb{V}$; see Table~\ref{tab:IEEEProInfo}.
In addition, for some records (i.e., pre-harvest sample, pre-harvest test, harvest, and certificate of analysis test after manufacturing), \textit{on-site investigation and verification} are required by the regulatory compliance to ensure that the corresponding critical data records (i.e., the THC content at pre-harvest test, harvest completed date, and THC level of final product) are accurate and reliable; see \href{https://oregon.public.law/rules/oar_chapter_603_division_48}{Division 48 Industrial Hemp}.

\begin{table*}[h]
\centering
\caption{IHSC participant information}
\label{tab:IEEEPartInfo}
\small
\begin{tabular}{lll}
Participant $\mathbb{P}$               & Information $\mathbb{I}$                                                                                                                                                                    & Verification  $\mathbb{V}$ \\ \hline
Breeder                       & Breeder Name,   Registration, PVP, Address                                                                                                                               & Online          \\ \hline
Licensed Grower        & \begin{tabular}[c]{@{}l@{}}Name, License, Field   Name, GPS, Background Check, \\ Pre-Plant Soil Tests, Irrigation   Type, \\ Prior Year Field History, Soil Type\end{tabular} & Online      \\ \hline
Dryer                  & Name, License, Address,   System Type                                                                                                                                          & Online      \\ \hline
Licensed Processor     & \begin{tabular}[c]{@{}l@{}}Processor-Extractor Name,   Address, \\ Handler License Information, System Type,\end{tabular}                                                      & Online      \\ \hline
Transporter            & Transporter Information, Driver License                                                                                                                                        & Online      \\ \hline
Hemp Testing Lab       & \begin{tabular}[c]{@{}l@{}}Lab Name, Address,   License, \\ And Other Related Information\end{tabular}                                                                         & Online      \\  \hline
Authorities/Regulators &Name, Address, License                                                                                                                                                         & Online  \\  
\end{tabular}
\end{table*}

\begin{table*}[h!]
\caption{IHSC process information}
\centering
\label{tab:IEEEProInfo}
\small
\begin{tabular}{llll}
Operation Unit $\mathbb{O}$                                                           & Signature $\mathbb{S}$                                                                       & Information  $\mathbb{I}$                                                                                                                                                                                                                      & Verification  $\mathbb{V}$ \\ \hline
Seed   Sourcing                                                              & \begin{tabular}[c]{@{}l@{}}Breeder, Transporter,\\ Grower\end{tabular}              & \begin{tabular}[c]{@{}l@{}}Variety, Seed   Lot   Number, Seed Purity Analysis, \\ Flowering Type, Feminization   Process, \\ Seed Feminization   Percentage, Clone Information,\\ Quantity\end{tabular}                               & Online            \\ \hline
Seed Pickup                                                                  & Breeder, Transporter                                                                & \begin{tabular}[c]{@{}l@{}}Transporter Information, Sender/ Receiver Information,\\ Pickup Date, Driver License\end{tabular}                                                                                                          & Online            \\ \hline
Seed Arrival                                                                 & \begin{tabular}[c]{@{}l@{}}Breeder, Transporter, \\ Grower\end{tabular}             & \begin{tabular}[c]{@{}l@{}}Shipment/Delivery Date,   Vehicle/Model Of Transport,  \\ Route of Transportation\end{tabular}                                                                                                             & Online            \\ \hline
\begin{tabular}[c]{@{}l@{}}Germination \&\\ Field   Preparation\end{tabular} & Grower                                                                              & \begin{tabular}[c]{@{}l@{}}Seeding/Transplanting   Date, Plant Density,   \\ Row Width,   Grown On Plastic, Lot Number,   GPS, \\ Pre-Plant Soil Test,  Irrigation    Type, Soil Type\end{tabular}                                    & Online            \\ \hline
Cultivation                                                                  & Grower                                                                              & \begin{tabular}[c]{@{}l@{}}Irrigation     Frequency/Volume, \\ Fertilizer Frequency/Volume,  \\ Weed/Insect/Mold/Pollination   Control\end{tabular}                                                                                   & Online            \\ \hline
Pre-Harvest Sample                                                           & Grower, Validator                                                                   & Pre-Harvest Hemp   Sampling and Testing Request Form                                                                                                                                                                                  &  Online           \\ \hline
Pre-Harvest Test                                                             & Lab                                                                                 & \begin{tabular}[c]{@{}l@{}}Sampling Date, Lot   Number, COA Test, \\ Cannabinoid Content, \\ Pesticide Residue,  Heavy Metals\end{tabular}                                                                                            & On-Site \& Online          \\ \hline
Harvest                                                                      & Grower, Validator                                                                   & \begin{tabular}[c]{@{}l@{}}Destruction or Harvest   Process And Completion Date, \\ Moisture Content,  Total Yield By Field\end{tabular}                                                                                              & On-Site \& Online       \\ \hline
IH Pickup                                                                    & \begin{tabular}[c]{@{}l@{}}Grower, Transporter,\\ Dryer\end{tabular}                & \begin{tabular}[c]{@{}l@{}}Transporter Information, Driver License, Lot Number,\\ Sender/Receiver Information, Quantity, Pickup Date\end{tabular}                                                                                     & Online            \\ \hline
IH Arrival                                                                   & \begin{tabular}[c]{@{}l@{}}Grower, Transporter, \\ Dryer\end{tabular}               & \begin{tabular}[c]{@{}l@{}}Shipment/Delivery Date,   Vehicle/Model of Transport,\\  Route of Transportation\end{tabular}                                                                                                              & Online            \\ \hline
Drying   Stabilizing                                                         & Dryer                                                                               & \begin{tabular}[c]{@{}l@{}}Lot Number, Pre-Drying   Weight, \\ Temperature/Duration of   Drying, \\ Final   Dried   Weight, Dry Weight, Container Type/Weight,\\  Cannabinoid Content, Pesticide   Residue, Heavy Metals\end{tabular} & Online            \\ \hline
Dried IH Pickup                                                              & \begin{tabular}[c]{@{}l@{}}Grower, Dryer,\\ Transporter, Processor\end{tabular}     & \begin{tabular}[c]{@{}l@{}}Transporter Information, Driver License, Lot Number,\\ Sender/Receiver Information, Quantity, Pickup Date\end{tabular}                                                                                     & Online            \\ \hline
Dried IH Arrival                                                             & \begin{tabular}[c]{@{}l@{}}Grower, Dryer, \\ Transporter,    Processor\end{tabular} & \begin{tabular}[c]{@{}l@{}}Shipment/Delivery Date,   Vehicle/Model of Transport, \\  Route of Transportation\end{tabular}                                                                                                             & Online            \\ \hline
Extraction                                                                   & Processor                                                                           & \begin{tabular}[c]{@{}l@{}}Lot Number, Biomass   Weight In, \\ Extraction Input   Quantity/Recaptured,  \\ Quantity Of Oil   Extracted, Gain/Loss Of Extraction, \\  Post Extraction Test\end{tabular}                                & Online            \\ \hline
Winterization                                                                & Processor                                                                           & \begin{tabular}[c]{@{}l@{}}Lot Number, Crude Oil   Weight In, Winterized Oil Out,\\ Post   Winterization Test\end{tabular}                                                                                                            & Online            \\ \hline
PLC                                                                          & Processor, Validator                                                                & \begin{tabular}[c]{@{}l@{}}Lot Number, Quantity/Weight   In, Quantity/Weight Out,\\ PLC   Repeat   Times,  Post PLC Test\end{tabular}                                                                                                 & On-Site \& Online         
\end{tabular}
\end{table*}

\subsection{Blockchain Design}
\label{subsec:blockchain}


Built on our previous studies \cite{zhouboson, wang2020simulationbased}, we develop the two-layer blockchain for the end-to-end IHSC. The blockchain is a distributed database, also a global ledger that records all the process data as a timestamp chain of blocks. 
For the existing IHSC design with only state regulators in charge of inspection, the verification efficiency totally depends on the number of available officials for on-site visit. Given the limited inspection resources at state and federal offices of USDA, it is challenging to handle the situation when the IH industry grows dramatically. \textit{Therefore, the two-layer blockchain platform allows us to leverage the additional resources from local authorities with state regulators, and improve the efficiency and security of IHSC.}

\begin{figure}[hbt!]
{
\centering
\includegraphics[width=0.45\textwidth]{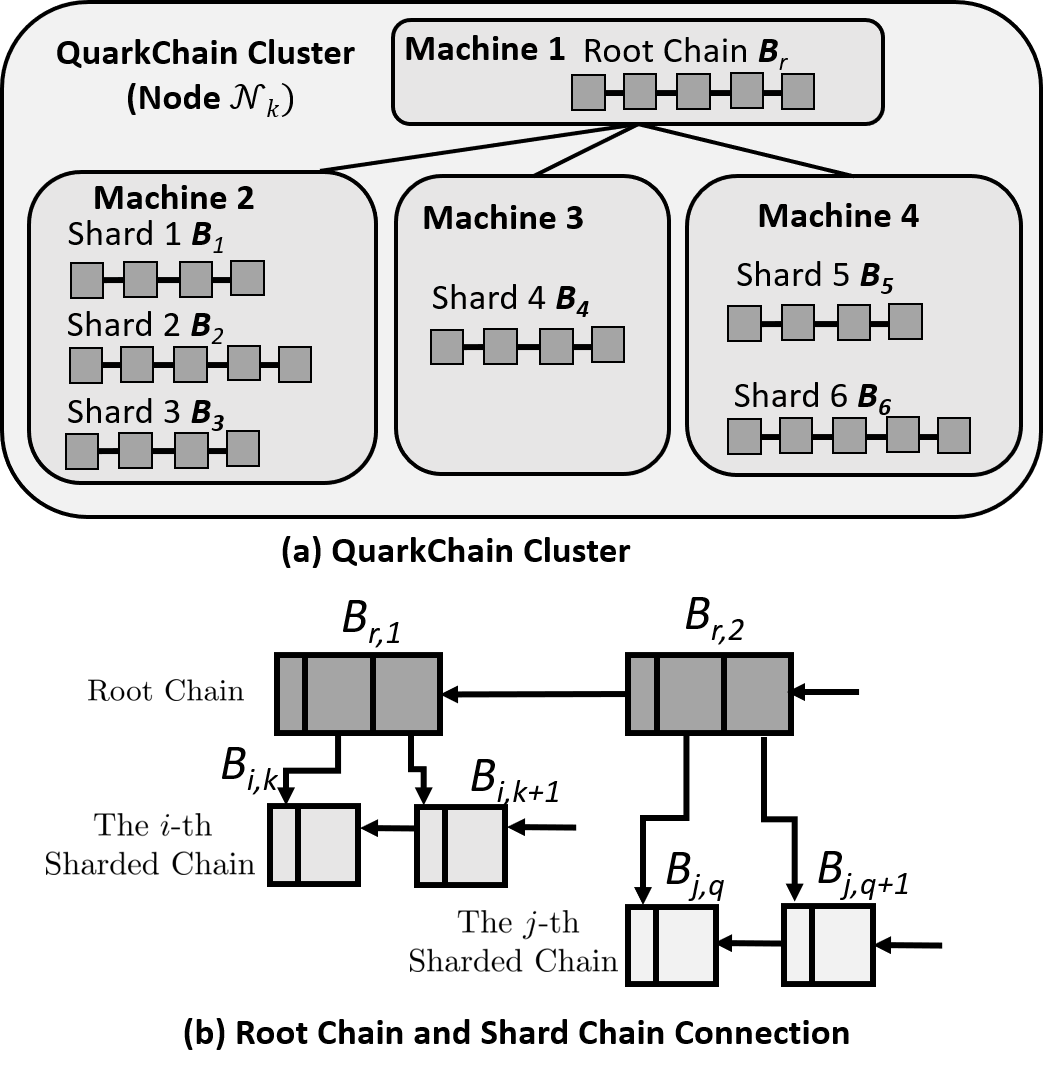}
\caption{The QuarkChain Cluster Illustration\label{fig: IEEEBCclu}}
}
\end{figure}

The two-layer blockchain includes: (1) the sharding layer supported by local authorities for on-site verification, which divides the IHSC into multiple areas or \textit{geography-based state sharding}; and (2) root chain layer supported by state regulators, which serves as the coordinator among all shard chains and has the final confirmation on the verified transactions and records. Figure \ref{fig: IEEEBCclu} illustrates the general  state-partition based two-layer blockchain design. 
Basically, a cluster is the equivalence of a “full node”, which keeps the full transaction history of the blockchain network. Here, a QuarkChain cluster, represents a node, say $\mathcal{N}_k$ with $k=1,\ldots,K$. It consists of a collection of machines/CPUs, one of which runs the root chain, and others run different shard chains; see Figure \ref{fig: IEEEBCclu}(a). Each block in a shard blockchain has two hash pointers: one links to the previous shard chain block and the other pointer links to a block in the root chain; see Figure \ref{fig: IEEEBCclu}(b).

\begin{figure}[]
{
\centering
\includegraphics[width=0.48\textwidth]{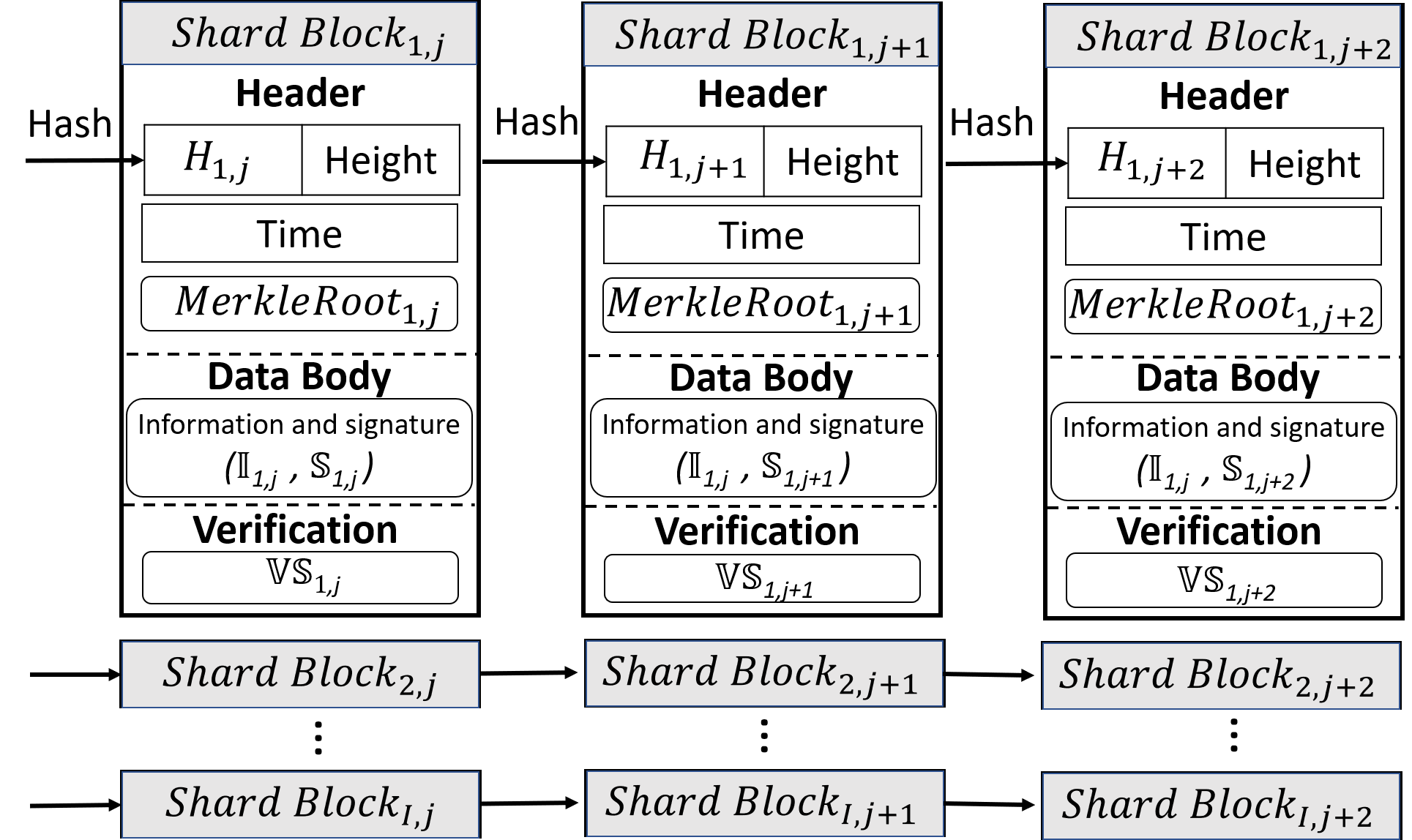}
\caption{Shard block structure design and n parallel shard chains  \label{fig: IEEEBCstruS}}
}
\end{figure}

Built on the participant information and transaction design presented in Tables~\ref{tab:IEEEPartInfo} and \ref{tab:IEEEProInfo}, the structure design of shard chains for IHSC is illustrated in Figure~\ref{fig: IEEEBCstruS}. 
\textit{For each $j$-th block in the $i$-th shard chain, denoted by $B_{i,j}$, there are three parts: Verification, Data Body and Header}. The Verification contains the signature of validator for on-site verification, 
denoted by $\mathbb{VS}_{i,j}$, if this record is needed by the regulation compliance based on Table~\ref{tab:IEEEProInfo}. 
The Data Body consists of critical data information, denoted by $\mathbb{I}_{i,j}$, and the signature of responsible participant(s), denoted by $\mathbb{S}_{i,j}$. The Header, denoted by $H_{i,j}$, returns: (1) the hash value of  previous block, $H_{i,j} = h(B_{i,j-1})$, (2) the height of current $i$-th shard chain, (3) the creating time of proposed block, and (4) Merkle root, denoted by $MerkleRoot_{i,j}$, 
 where $h(\cdot)$ is a cryptographic hash function.
The Merckle root is the hash of Data Body and Verification (if there is any required on-site verification): 
\begin{equation}
MerkleRoot_{i,j} = h(\mathbb{I}_{i,j},\mathbb{S}_{i,j},\mathbb{VS}_{i,j}). \nonumber 
\end{equation}

\begin{figure}[hbt!]
{
\centering
\includegraphics[width=0.48\textwidth]{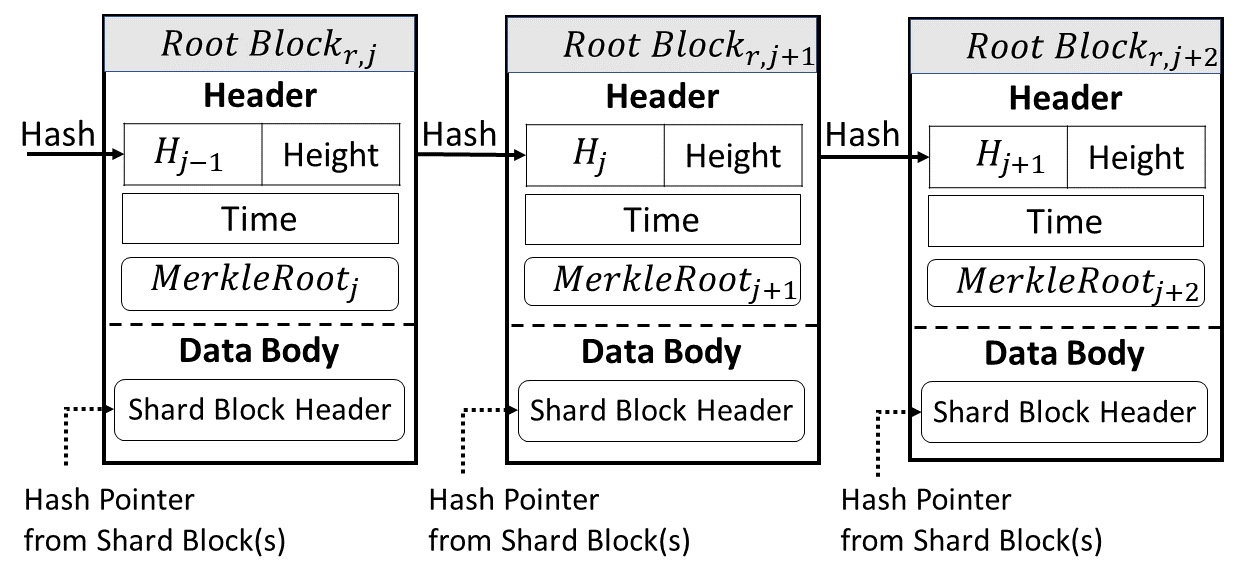}
\caption{The root chain structure design\label{fig: IEEEBCstruR}}
}
\end{figure}

The structure design of the root chain for IHSC is illustrated in Figure~\ref{fig: IEEEBCstruR}.
\textit{For each $j$-th block in root chain $r$, denoted by $B_{r,j}$, it has three parts: Confirmation, Data Body and Header.} 
The confirmation contains the signature of regulatory, denoted by $\mathbb{CS}_{r,j}$, who confirms the verified information contained in the corresponding shard block. 
The root block describes the canonical chain of each shard chain by including the hash pointers or headers of the last shard block observed,
\begin{equation}
    B_{i,o_i(B_{r,j})} = ob_i(B_{r,j}), \nonumber
\end{equation}
\noindent where $o_i(B_{r,j})$ means the highest index of the block of the $i$-th shard chain included 
until the root block $B_{r,j}$; see the illustration example in Figure~\ref{fig: IEEEExample}.

The Data Body describes the canonical chain of each shard chain by including the hash pointer or header of the shard block $B_{i,o_i(B_{r,j})}$.
The Header of root chain 
includes the hash of Data Body and Confirmation. 
\textit{Therefore, following this unique mechanism of each root and shard block, a two-layer blockchain is created for IHSC to improve the safety/efficiency/throughput and support the scalability and data integrity.} 

\textbf{Remark:} The Verification in the shard block and the Confirmation in the root block are only required for those transactions that need to follow the regulation specifications and compliance; see Table~\ref{tab:IEEEProInfo}.

\vspace{-0.1in}

\subsection{Blockchain Network, Smart Contract, and Consensus Design for Regulated IHSC}
\label{subsec:consensus}

Based on the unique features of IHSC, in this section, we discuss blockchain network and design smart contract and consensus, including hierarchical verification and validation for two-layer blockchain. For the highly regulated IH industry, Proof-of-Authority (PoA) based smart contract, denoted by $\mathbb{SC}$, is designed to ensure that the blockchain contains valid and reliable data and information. It can support regulatory compliance and quality verification,
control data tampering and cyberattacks, and improve the IHSC safety and data integrity.

The two-layer blockchain provides the ledger recording all historical data and transactions occurring in the IHSC, represented by 
 \begin{equation}
   \mathbb{L} \equiv (\textbf{B}_r,\textbf{B}_1,\textbf{B}_2,\dots,\textbf{B}_I),
   \nonumber
\end{equation}
where $\textbf{B}_r = [B_{r,0},B_{r,1},\dots,B_{r,l_r}]$ is the list of blocks included in the root chain, $I$ is the number of shard chains, and $\textbf{B}_i = [B_{i,0},B_{i,1},\dots,B_{i,l_i}]$ represents the list of blocks recorded in the $i$-th shard chain for $1\leq i \leq I$.
The ledger $\mathbb{L}$, providing the world state of data that have been recorded in blockchain, is replicated across different nodes, and it is shared among the internet-connected participants of the IHSC.

Suppose that each $k$-th node, denoted by $\mathcal{N}_k$, corresponds the $k$-th participant $N_k$ with a single device. 
All nodes are equally involved in maintaining the blockchain, but different nodes can have different roles for running the blockchain. In this paper, we consider 
a fixed set of nodes. Let $
\mathbf{N} \equiv (N_1,N_2,\dots,N_K)$ denote all participants, where $K$ is the number of participants. Since the \textit{validated blockchain} is replicated and broadcast among the nodes through the internet network, we represent the blockchain network, 
\begin{equation}
    \mathcal{{N}} \equiv \{\mathcal{N}_1(\mathbb{L},\mathbb{SC}), \mathcal{N}_2(\mathbb{L},\mathbb{SC}),\dots ,\mathcal{N}_K(\mathbb{L},\mathbb{SC})\} \nonumber
\end{equation}
where $\mathcal{N}_k(\mathbb{L},\mathbb{SC})$ for $k = 1,2,\dots,K$ represents Node $k$ with blockchain ledger copy $\mathbb{L}$ and PoA based smart contract, denoted by $\mathbb{SC}$, as a piece of code residing on this blockchain.

\textit{Each new record and/or transaction will be processed by smart contract 
for validation, the validated shard and root blocks will be added to the two-layer blockchain, and then the updated blockchain will be broadcast to the blockchain network.} To facilitate the interoperability and support the participants reaching a common global view of the world state, each blockchain network needs to establish consensus rules that each data record (or transaction) should conform to. The consensus $\mathbb{C}$ contains two elements: 
the validity function $V(\cdot)$, and fork-choice rule function. 
Here, we are mainly focusing on the \textit{validity function}. The proposed validation for IHSC blockchain is built on the Boson consensus introduced in \cite{zhouboson}, which can scale the network dramatically while keeping security guaranteed.

\textit{We first discuss the validation for shard chain.} Each new $j$-th shard block $B_{i,j}$ at the $i$-th shard chain with  $i=1,2,\ldots,I$ includes multiple transactions and/or records. For any new information $\mathbb{I}$ coming to shard chain $i$, if it is required by the regulation compliance, 
the smart contract  will send out a verification request to the local authority which will then dispatch an investigator for on-site verification,
\begin{equation}
v_i^a(\mathbb{I}) = \left \{
\begin{array}{rl}
1,  & \text{If $\mathbb{I}$ is approved by a randomly} \\
    & \text{selected local investigator from area $i$;} \\
0,  & \text{otherwise.} \label{IEEEeqV}
\end{array} \right. 
\end{equation}
For those transactions $\mathbb{I}$ that only require online verification, we set $v_i^a(\mathbb{I})=1$. Each shard block has the size limit and blocks are generated per block interval. The verified messages will be uploaded to the block $B_{i,j}$ until reaching the block size. Thus, the block $B_{i,j}$ is called verified, represented by $V_i^a(B_{i,j}) = 1$, which means each transaction or record included in this block is verified, $v_i^a(\mathbb{I})=1$ for any $\mathbb{I} \in B_{i,j}$.

Then, the \textit{validation function} for the $i$-th shard chain becomes, 
\begin{equation}
 V_i(B_{i,j}) = \left\{
\begin{array}{rl}
1,       & V_i^a(B_{i,j}) = 1, \\ 
         & V_i(B_{i,j-1})= 1, ~\text{and}\\
         & h(B_{i,j-1}) = pre\_hash(B_{i,j}); \\
0,       & \text{otherwise,} \label{IEEEeqVSnV}
\end{array} \right. 
\end{equation}
for any $i=1,2,\ldots,I$,
where $pre\_hash(B_{i,j})$ returns the hash value of the previous block, $h(B_{i,j-1}) = pre\_hash(B_{i,j})$. The new shard block $B_{i,j}$ is validated if the previous block is already validated and it is also hash-linked to $B_{i,j-1}$. 
Then, the verified data together with investigator’s signature will be included in a new shard block. 

\begin{figure}[hbt!]
{
\centering
\includegraphics[width=0.48\textwidth]{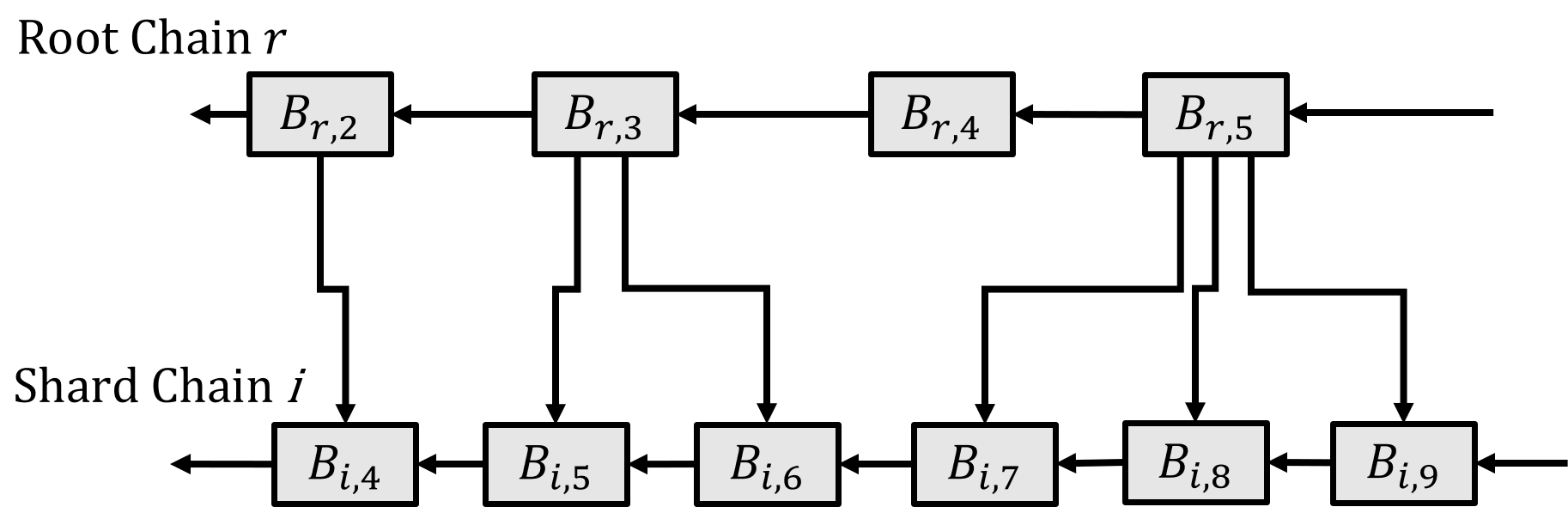}
\caption{The simple illustration example of the root chain structure design\label{fig: IEEEExample}}
}
\end{figure}

\begin{sloppypar}
\textit{After that, we discuss the validation function for each root block.}
At the root chain, 
a regulatory official simply confirms the record/information included by each shard block. At this point, we suppose the regulatory officials completely trust the decision made by local authorities. Each new root block, denoted by $B_{r,q+1}$, includes the headers of multiple new shard blocks (see an illustrative example in Figure~\ref{fig: IEEEExample}), 
where $q$ represents the highest index of the block included in the root chain. Denote the highest index of the root block which contains the $i$-th shard chain's block as $h(i)$. Similar to shard chains, each root block has a size limit and blocks are generated based on certain interval. 
Then, the validation for the new root block is,
\begin{align}
V_r(B_{r,q+1}) = \left\{
\begin{array}{rl}
1,       
         & V_r(B_{r,q}) = 1,\\
         & h(B_{r,q}) = pre\_hash(B_{r,q+1}),\\
         & ob_i(B_{r,q+1}) \xrightarrow{o_i(B_{r,q+1})-o_i(B_{r,h(i)})} \\     
         &  ob_i(B_{r,h(i)}), ~\text{and} \\
         & V_i(B_{i,k}) = 1 ~\text{for} \\
         & o_i(B_{r,h(i)})+1 \leq k \leq o_i(B_{r,q+1}).\\
0,       & \text{otherwise,} \label{IEEEeqVRnC}
\end{array} \right.
\end{align}
for any $i=1,2,\ldots,I
$,
where 
\begin{equation}
ob_i(B_{r,q+1}) \xrightarrow{o_i(B_{r,q+1})-o_i(B_{r,h(i)})}ob_i(B_{r,h(i)})
\label{IEEEeqVRnC:link}
\end{equation}
represents the block $ob_i(B_{r,q+1})$ and $ob_i(B_{r,h(i)})$ are $\left[o_i(B_{r,q+1})-o_i(B_{r,h(i)})\right]$-hash-linked; 
see the detailed description in \cite{zhouboson}.
In addition, all shard blocks $B_{i,k}$ with index $o_i(B_{r,q})+1 \leq k \leq o_i(B_{r,q+1})$ and the previous root block $B_{r,q}$ 
should be validated.
\textit{Once the new root block is validated, this root block and its corresponding shard blocks are generated and added to the blockchain.}
\end{sloppypar}

To make it easy to follow, we use the simple example in Figure~\ref{fig: IEEEExample} to illustrate the implementation of hash-link in (\ref{IEEEeqVRnC:link}), 
and the value assigned for each element.
Here, we consider three representative situations. First, both previous and appending root blocks 
contain the headers of shard blocks from $i$-th shard chain (i.e., $B_{r,2}$ and $B_{r,3}$). We have $ob_i(B_{r,q+1}) = ob_i(B_{r,3}) = B_{i,6}$, $ob_i(B_{r,h(i)}) = ob_i(B_{r,2}) = B_{i,4}$, $o_i(B_{r,q+1})-o_i(B_{r,h(i)}) = 6-4 =2$, and $B_{i,6} \xrightarrow{2}B_{i,4}$, which means the $B_{i,6}$ and $B_{i,4}$ are 2-hash-linked.  
Second, the previous root block has but the appending one doesn't have the headers from the $i$-th shard chain (i.e., $B_{r,3}$ and $B_{r,4}$). Then,
$ob_i(B_{r,q+1}) = ob_i(B_{r,4})$, which is not further updated from $ob_i(B_{r,3})$. Thus,
$B_{i,6} \xrightarrow{0}B_{i,6}$, and the block $B_{i,6}$ has been validated through Equation~(\ref{IEEEeqVSnV}).
Third, the previous root block doesn't have but the appending one has the headers from the $i$-th shard chain (i.e., $B_{r,4}$ and $B_{r,5}$).
We have $ob_i(B_{r,q+1}) = ob_i(B_{r,5}) = B_{i,9}$, $ob_i(B_{r,h(i)}) = ob_i(B_{r,3}) = B_{i,6}$, $o_i(B_{r,q+1})-o_i(B_{r,h(i)}) = 9-6 =3$, and the $B_{i,9}$ and $B_{i,6}$ are 3-hash-linked as $B_{i,9} \xrightarrow{3}B_{i,6}$. 

For the fork-choice rule function, when new IHSC data/ information records need to be written in the global ledger, new blocks will be generated and added to the blockchain as long as it is defined validly by $V(\cdot)$, and broadcast to the network so that each node will update the copy of the ledger. However, if another block is produced at the same height and results in different ledgers, namely, forks, the network reaches an inconsistent state temporarily, then the fork-choice rule will determine which fork to survive. The fork-choice rule is beyond paper's scope, more detailed descriptions are presented at \cite{zhouboson}.

\section{Blockchain-Enabled IoT Platform Development for IHSC Management}
\label{sec:implementation}
 
We describe the 
architecture of the proposed blockchain-enabled IoT platform in Section~\ref{subsec:blockchainFramework}, which can be extended and generalized to various application areas, such as regulated  biopharmaceutical manufacturing and supply chain. 
After that, we describe how various messages flow through the cyber network of IHSC and provide the algorithm design for the platform in Section~\ref{subsec:algorithm}.

\subsection{Architecture of Proposed Platform}
\label{subsec:blockchainFramework}

\begin{figure*}[htb]
{
\centering
\includegraphics[width=0.85\textwidth]{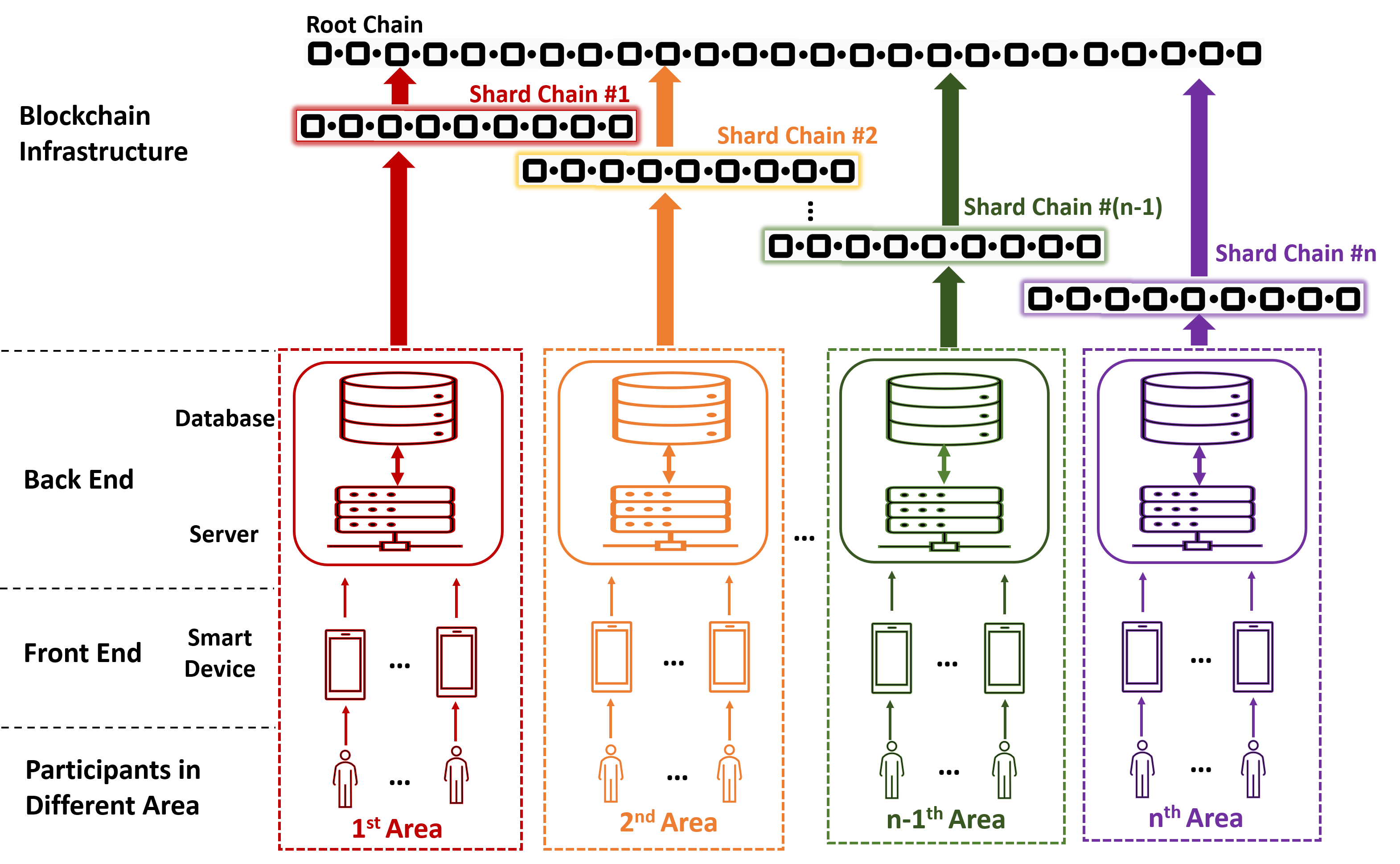}
\caption{Architecture of proposed Blockchain-Enable IoT Framework\label{fig: IEEEArchi}}
}
\end{figure*}

\textit{The proposed blockchain-enabled IoT platform is composed of the key components, including: front-end mobile user interface (UI), back-end server, database, and blockchain Infrastructure; see the illustrative plot in Figure~\ref{fig: IEEEArchi}.}
More specially, it consists of: (1) cross-platform mobile application for the front end, which supports both iOS and Android operating system; 
(2) Node.js project for back-end server; (3) MongoDB for NoSQL database; and (4) QuackChain clusters for blockchain infrastructure.

We create the geography-based state partition and blockchain sharding in the proposed platform so that it can simultaneously process the jobs coming from different areas, support the interoperability, and improve the efficiency and throughput. 
Basically, the records from $i$-th area are processed and verified by the $i$-th shard chain, and the root chain coordinates different shard chains.

\underline{\textbf{Mobile User Interface:}} 
\textit{The users of IHSC blockchain-enabled IoT platform can be divided into two categories: record creator and data/information visitor.} The record creators include those participants who need to write data and information into the blockchain system, such as 
seed companies, farmers, transportation companies, hemp testing lab, drying companies, and manufacturers. They can write the records into the platform through unloading the information by using our developed cross-platform (Android and iOS) mobile application. The uploaded information will be further validated, and only validated messages can be saved in the blockchain system. 

The record visitors include those participants who want to track the detailed historical information about certain products. Their primary use of this platform is to check the record on the IH sourcing, cultivation, and processing for various safety and efficiency related needs, such as regulatory compliance monitoring, etc. For each IH final or semi-manufacturing product, there is a unique product ID, denoted as $P^{id}$. 
\textit{Thus, the visitor can directly retrieve the product historical record through the mobile app and blockchain system by utilizing the product ID.}    



\underline{\textbf{Back-End Server:}} We provide a back-end service to manage the data flow from Mobile User Interface (UI) to Database and Blockchain. Mobile applications can send requests to back-end servers through RESTful Application Programming Interface (API). Currently, there are about 50 APIs developed in our platform, which automatically support user create, user login, companies create, companies modify, product create, processing information adding, transportation information adding, etc. Back-end servers are connected with the two-layer blockchain clusters by using Quarkchain-web3.js. Since blockchain runs Ethereum Virtual Machine, this library is built on top of web3.js and supports smart contracts. We design and deploy smart contracts on blockchain mainnet, which is used to automatically manage the data and control the dynamic interactions of the entities or participants involved in the IHSC system, such as users, companies, products, and transportation packages. Quarkchain-web3.js can send JSON RPC calls to the mainnet so the back-end servers can interact with the smart contracts in the blockchain infrastructure automatically. 

\underline{\textbf{Database:}} MongoDB is used as a backup to store user information and large data sets. For any large file, such as pictures/videos of IH harvesting process, and PDF files of the comprehensive seed cloning and hemp testing results, 
we first compute the SHA-256 hash of the file and save the hashed string into blockchain. The actual file is saved in the MongoDB. When a user wants to retrieve these files, they will download them from the back-end server and compare the SHA-256 signature with the information/data saved in the blockchain. Since the large-size files are not saved in blockchain, this procedure can help to ensure the data integrity. Additionally, in order to counter the malicious cyberattacks, we deploy several back-end servers and databases as backup. 

\underline{\textbf{Blockchain Infrastructure:}} In the proposed platform, we develop the two-layer blockchain with Proof-of-authority based hierarchical smart contract; see the the structure description in Section~\ref{subsec:blockchain}. Moreover, due to the state partition and sharding technique, the blockchain infrastructure utilizes the parallel computing and processing. It has the excellent performance in terms of scalability and speeding up verification and validation. The records/transactions/jobs coming from different areas can be processed simultaneously, which can support interoperability, and greatly improve the IHSC safety and efficiency. 

\subsection{Algorithm design}
\label{subsec:algorithm}

Based on the blockchain-enabled IoT platform and its architecture described in section~\ref{subsec:blockchainFramework}, we present the overall flowchart and algorithms of record validation and information/data retrieve. We describe how the messages flow through the proposed platform and present the algorithm implementation.


\begin{figure*}[htb]
{
\centering
\includegraphics[width=1\textwidth]{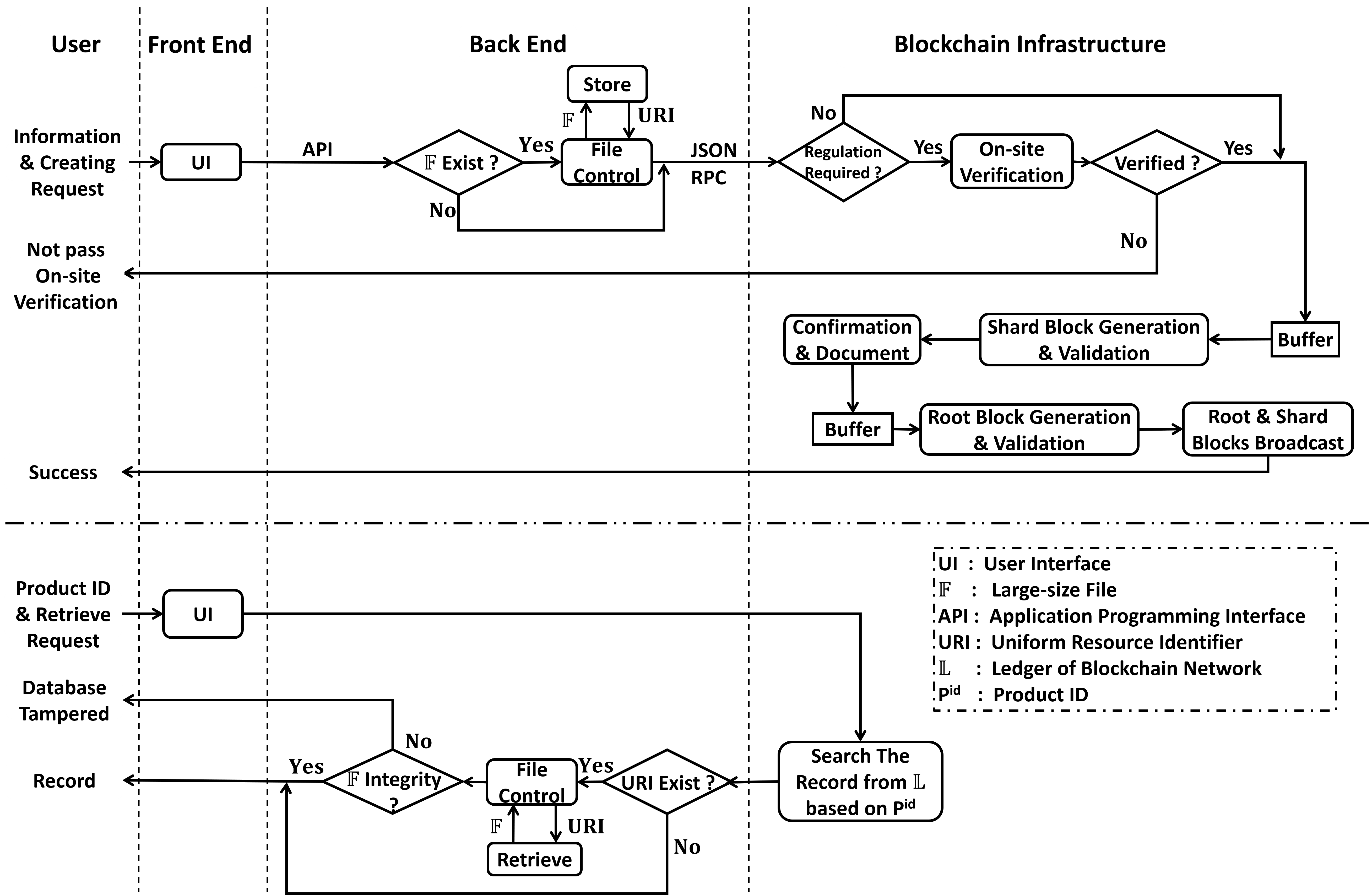}
\caption{Message flow of proposed Blockchain-Enable IoT Framework \label{fig: IEEEOverall}}
}
\end{figure*}

\textit{The implementation procedure of record creation
is described in Algorithm~\ref{IEEEalgupload}.} The record creators write the validated record into the blockchain-enabled IoT system with main steps and message flow illustrated in the top part of Figure ~\ref{fig: IEEEOverall}. Specifically, the creators input crucial IHSC information/data into our developed mobile app, which will be sent to back-end server through Restful API.
If the data contain the large-size file (i.e., picture and/or video), denoted by $\mathbb{F}$, the server will store it in the database, obtain the Uniform Resource Identifier (URI), and compute its hash. Then, the server sends the URI and hash string to the blockchain mainnet along with other digital/textual information through JSON RPC. Once receiving the message, if this instance is relative to the IHSC regulatory compliance, the smart contract on shard chain will randomly select one available local authority and dispatch him to have on-site investigation. If it passes the verification, the record waits at buffer for the generation of next shard chain block. Otherwise, the error message showing the verification failure will send back to the user.

At any shard block generation time, the local authority, represented by one node, constructs a shard block containing the new verified records,
and attempt to append the block to shard chain validly, based on Equation~(\ref{IEEEeqVSnV}). 
The header of the newly generated shard block will be sent to root chain and wait for validation and next root block generation. A randomly selected regulatory official will confirm and validate the new root block according to Equation~(\ref{IEEEeqVRnC}). After validation, the newly generated shard and root blocks will be broadcast to the whole blockchain network. 
Then, the success message will return to user.

\begin{algorithm*}
\DontPrintSemicolon
\KwIn{IHSC Information/Data $\mathbb{I}_i= \{\mathbb{D}_i,\mathbb{F}_i\}$, where  $\mathbb{D}$ is the digital/textual information and $\mathbb{F}$ represents the large-size file (i.e., picture and video), signature of corresponding participants $\mathbb{S}_i$, current ledger of blockchain network $\mathbb{L}$. Suppose the current heights of $i$-th shard chain and root chain are $j-1$ and $q-1$ respectively.}
\KwOut{
Return the status of record creation on the blockchain: $U(\mathbb{I}_i,\mathbb{S}_i,\mathbb{L})=1$ for success; and 0 for failure.}
\SetKwBlock{Begin}{Function:}{end function}
\Begin(\text{Create new record of data/transaction, conduct the validation, and update the blockchain network.}
)  
{
    \textbf{Step (1)} User inputs the corresponding information $\mathbb{I}_i$ and signature $\mathbb{S}_i$ through the front-end mobile app. \\
    \textbf{Step (2)} The front end sends the record and request to the back-end server through pre-designed RESTful API. \\
    \textbf{Step (3)} Pre-process the large-size file at the back end if needed. \\
    Step (3.1) Large-size file control: \\
    \uIf{there is large-size file $\mathbb{F}_i$}
    {
    Store $\mathbb{F}_i$ at the database and compute the hash of file $h(\mathbb{F}_i)$. \\
    }{\textbf{end}} \\
    Step (3.2) The server sends the collection of all information $\mathbb{R}_i = \{\mathbb{D}_i, \mathbb{S}_i, h(\mathbb{F}_i) , \text{URI of large-size file}\}$ to blockchain smart contract through JSON RPC; \\
    \vspace{0.03in}
    \textbf{Step (4)} Smart contract for on-site verification control: \\
    \uIf{$\mathbb{I}$ is relative to regulation}
    {
    Once receiving, the smart contract $\mathbb{SC}_i$ on shard chain $i$ will randomly dispatch a local authority in charge of area $i$ to investigate on-site. \\
    Update $v^a_i$ based on Equation ~(\ref{IEEEeqV}).\\
    \uIf{ $v^a_i = 0$}{\textbf{Return} Message ``unapproved by local authority" to user and terminate the whole process.}{\textbf{end}} \\
    }{\textbf{end}} \\
    \vspace{0.03in}
    \textbf{Step (5)} Shard block validation and generation.\\
    Step (5.1) After on-site verification, the record $\mathbb{R}_i$ needs to wait at buffer for the next block generation. 
     \\  
    Step (5.2) At next shard block generation time, the authorized node generates the shard block $B_{i,j}$, which contains the record $\mathbb{R}_i$ and other records from same area $i$, based on the structure introduced at Figure~\ref{fig: IEEEBCstruS}.\\
    Step (5.3) The node attempts to append the shard block to shard chain $i$ validly based on Equation ~(\ref{IEEEeqVSnV}).\\
    Step (5.4) Once validated, the new shard block $B_{i,j}$ is connected to shard chain $i$.\\
    \vspace{0.03in}
    \textbf{Step (6)} Root block validation and generation.\\
    Step (6.1) The shard-block header $B_{i,j}.Header$ is sent to the root chain. The regulatory official confirm it online. 
    \\
    Step (6.2) After confirmation, the header $B_{i,j}.Header$ waits at buffer for being written to root chain.
    \\ 
    Step (6.3) At next root block generation time, the regulatory official node generates the root block $B_{i,q}$, containing $B_{i,j}.Header$ and other headers from different shard chain, based on the structure introduced at Figure~\ref{fig: IEEEBCstruR}.\\
    Step (6.4) The node attempts to append the block to root chain validly based on Equation ~(\ref{IEEEeqVRnC}). \\
    Step (6.5) If validated, the new root block $B_{i,q}$ is appended to root chain.\\
    \vspace{0.03in}
    \textbf{Step 7)} New blocks' broadcast and ledger updating.\\
    Step (7.1) The new root block and corresponding shard blocks are broadcast to the whole blockchain network.\\
    Step (7.2) All node receive them and update their ledger $\mathbb{L}$.\\
    Step (7.3) Update $U(\mathbb{I}_i,\mathbb{S}_i,\mathbb{L}) = V_r(B_{r,q})$. \\ 
    \textbf{Return} Message to user based on $U(\mathbb{I}_i,\mathbb{S}_i,\mathbb{L})$.
}
\caption{Two-Layer Blockchain-Enabled IoT Platform for Process Quality Validation and Record Creation}
\label{IEEEalgupload}
\end{algorithm*}

\textit{The implementation procedure of information or record retrieve is
illustrated in Algorithm~\ref{IEEEalgRetrieve}.} 
Given the product ID, the visitor can retrieve its historical records along the end-to-end IHSC. 
The main steps of message flow is shown in the bottom part of Figure ~\ref{fig: IEEEOverall}. The visitor can simply send the retrieve request with product ID $P^{id}$ to blockchain through the Mobile App and back-end server.  
Once obtaining the request, the corresponding smart contract on blockchain will obtain all historical records based on $P^{id}$ from ledger $\mathbb{L}$. If the URI exists, the server will use it to retrieve the large-size file $\mathbb{F}$ from Database, and compare the hash value of obtained file $h(\mathbb{F})$ with the hash string from blockchain to check the integrity of $\mathbb{F}$. If $\mathbb{F}$ matches with the corresponding blockchain record, all retrieved records will be sent back to the visitor's Mobile App through well-designed hierarchical information check pages. Otherwise, the error message, ``database is tempered," will return.

\begin{algorithm*} 
\DontPrintSemicolon
\KwIn{Product ID $P^{id}$, current state/ledger of blockchain network $\mathbb{L}$.}
\KwOut{Historical records of this product.}
\SetKwBlock{Begin}{Function:}{end Function:}
\Begin($\text{Retrieve IHSC Records.}
$) 
{   
    \textbf{Step (1)} User provides the product ID $P^{id}$ at the front-end mobile app.\\
    \textbf{Step (2)} The front end sends requests and $P^{id}$ to the smart contract on blockchain through the backend server. \\
    \textbf{Step (3)} Once receiving the request, the smart contract searches and gets the historical records from ledger $\mathbb{L}$ based on $P^{id}$.\\  
    \textbf{Step (4)}: Large-size file retrieve and validation at the back end. \\
    Step (4.1) Large-size file control: \\
    \uIf{URI exsits}{Based on the URI, obtain the large-size file from the database.}{\textbf{end}}\\
    Step (4.2) Computes the hash value of the file and compares it to the hash value retrieved from the blockchain to validate the file’s integrity;\\
    \uIf{Doesn't match}
    {
        \textbf{Return} Message ``error, database is tampered" to the user and terminate the whole process.\\
    }{\textbf{end}}\\
    \textbf{Return} Historical records to the user.\\
}
\caption{Two-layer Blockchain-Enabled IoT Platform for Record Retrieve}
\label{IEEEalgRetrieve}
\end{algorithm*}

\section{Performance Analysis}
\label{sec:performance}

The developed blockchain-enabled IHSC platform is tested and validated during the small-scale pilot phase: industrial hemp season 2020 
in Oregon, Alabama, Colorado, and Pennsylvania. 
In this section, we conduct the virtual experiments to study the performance of this platform in terms of improving safety, efficiency, transparency, and process quality control. 

According  to \cite{sterns2019emerging}, the total acres licensed for IH production is 310,721 in 2019 nationally, and the average acres per lot is 8--10. Thus, we set the number of lots to be $K=320,000/8 =40,000$. 
In addition, we set: (1) the number of transplant and harvest machines, $n_f = 8,000$; (2) the number of pre-harvest test equipment, $n_l = 8,000$; (3) the number of drying machines, $n_d = 2,400$; (4) the number of machines used for extraction, winterization and PLC, $n_p = 1,600$. To avoid the product destruction required by the regulation, suppose each participant could tamper data with probability $p_2 = 30\%$. For the two-layer blockchain, let the verification and confirmation times follow the exponential distributions, i.e., $F_v \sim \exp(\mu_v)$ and $F_{c} \sim \exp(\mu_{c})$ with means $\mu_v=0.1$ and $\mu_{c}=0.005$. The generation interval of shard block and root block are assigned as 15s and 1.5min. 
The blockchain has 4 shard chains. There are $n_s=175$ local validators assigned to each shard chain and $n_r = 50$ state regulators in charge of confirmation at the root chain.
Assume each shard block can contain the information up to 4 transactions and each root block can have up to 24 transaction headers. Other setting parameters are assigned based on \cite{wang2020simulationbased}. 

\textit{We first show that the blockchain-enabled IHSC can improve the safety.} Here we compare the performance of IHSC with and without blockchain. For the critical record (see Table~\ref{tab:IEEEProInfo}), the validators need to have on-site visits to ensure the \textit{data integrity}. To assess the IHSC safety, we consider three types of false pass rates, including: (1) for pre-harvest test, the expected percentage of lots with more than 0.3\% THC that aren't destroyed, $q_{fp} = \underset{K\to\infty}{\lim}\mbox{E}[K_{fp}/K]$; (2) for the 15-day harvest regulation, the expected percentage of lots completed harvest violating the 15 days requirement, $q_{fh} = 
\underset{K\to\infty}{\lim}\mbox{E}[K_{fh}/K]$; and (3) the expected percentage of lots with more than 0.05\% THC and reaching to customers, $q_{fq} =  \underset{K\to\infty}{\lim}\mbox{E}[K_{fq}/K]$, where $K_{fp}$, $K_{fh}$, $K_{fq}$ are the counts of corresponding false approval. The simulation results in Table~\ref{IEEESecuImpor} illustrate that blockchain 
can greatly improve \textit{IHSC security and safety}.

\begin{table}[hbt!]
\centering
\caption{Simulation results of security improvement.}
\label{IEEESecuImpor}
\begin{tabular}{|c|c|c|}
\hline
{Security}               & {With Blockchain} & {Without Blockchain}  \\ \hline
{False Pass Pre-Harvest} & {0$\pm$0\%} & {2.32$\pm$1.04\%} \\ \hline
{False Pass Harvest}     & {0$\pm$0\%} & {3.13$\pm$2.56\%}  \\ \hline
{Fake Qualified}         & {0$\pm$0\%} & {0.65$\pm$0.43\%}  \\ \hline
\end{tabular}
\end{table}

{Then,
we compare the performance of a traditional single chain with our two-layer block chain with 4 shards.} For the single chain, suppose that the verification time for each record follows the exponential distribution, $F_s \sim \exp(\mu_s) $ with mean $\mu_s = 0.1$ day. The block size is 4 transactions and the generation interval is 1.5 min. 

\begin{figure}[h!]
{
\centering
\includegraphics[width=0.48\textwidth]{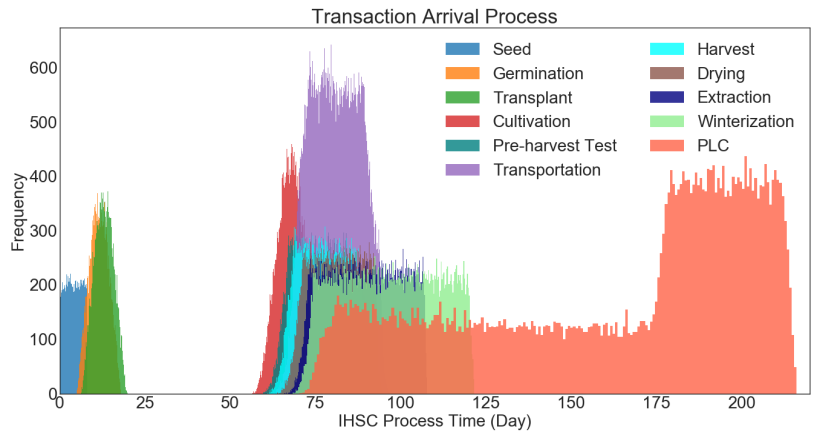}
\caption{Transaction arrival along the IHSC process \label{fig: IEEEArrival}}
}
\end{figure}

The frequency of all types of record arrivals along the IHSC process during one season is presented at Figure~\ref{fig: IEEEArrival}(a). 
The important percentiles (95\%, 50\%, and 5\%) of the corresponding waiting times for  online validation obtained by single-chain v.s. our two-layer blockchain are showed at Figures~\ref{fig: IEEEArrival}(b) and \ref{fig: IEEEArrival}(c) respectively. Moreover, Figures~\ref{fig: IEEEArrival}(d) and \ref{fig: IEEEArrival}(e) illustrate the results of waiting times for on-site verification. \textit{The plots show that the two-layer blockchain structure can improve the efficiency and throughput of both online validation and on-site verification, support scalability, and avoid the prolonged waiting.} 

\begin{figure*}[h]
{
\centering
\includegraphics[width=1\textwidth]{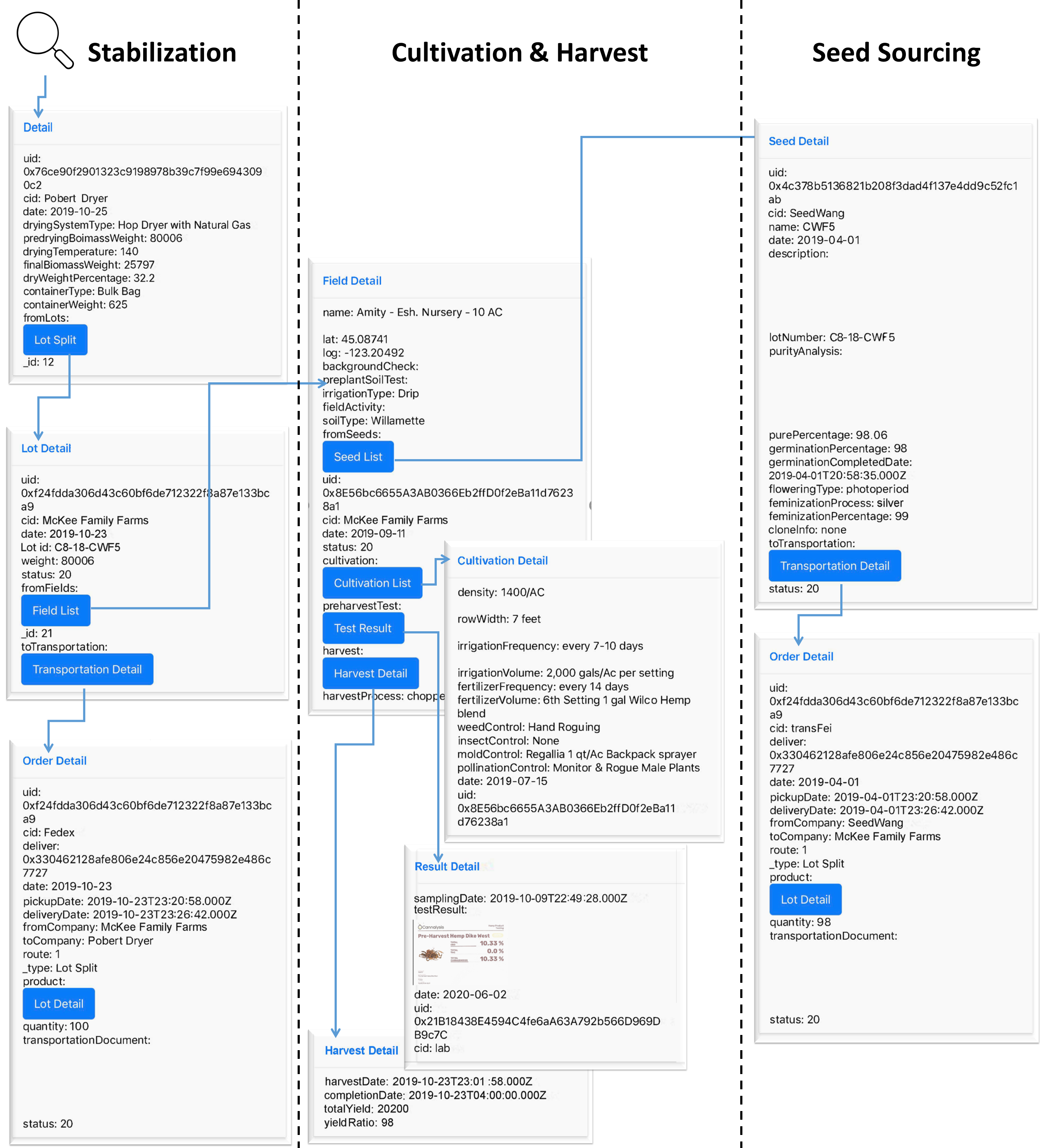}
\caption{Historical Information and Record Retrieve through the Mobile App  \label{fig: IEEEImple}}
}
\end{figure*}

Finally, we validate and implement the proposed blockchain-enabled IoT platform by using the real-world IHSC data obtained in Oregon. After the data are validated and written into the blockchain, \textit{the user can trace back all historical information on any product along the IHSC.} For each IH final and semi-manufacturing product, we can use the product ID to track and retrieve the historical data and information 
by using the developed mobile app and blockchain; 
see an example illustration in Figure~\ref{fig: IEEEImple}. By easily click the information card (blue block), the user can obtain all historical records, including seed, cultivation, pre-harvest test, harvest, drying, and transportation.

\section{Conclusion}
\label{sec:conclusion}
For the end-to-end regulated industrial hemp supply chain (IHSC), we create the blockchain-enabled internet-of-things (IoT) platform to improve the transparency, interoperability, safety, security,  traceability, and throughput. We 
propose geography-based state partition, two-layer blockchain design, hierarchical proof-of-authority based smart contract and consensus design. We further develop a user-friendly mobile app so that each participant can use internet-connected smartphones and devices to real-time collect and upload the validated data to the blockchain system, and track the historical data and industry hemp production and shipment along the integrated supply chain process. Thus, the proposed platform can support the interoperability, accelerate the product quality control and validation, and facilitate the dynamic information/data tracking.
The empirical study indicates that the proposed blockchain-enabled IoT platform has promising performance, and the description of practical implementation is provided. Our platform can be extended and utilized in various application areas, such as highly regulated global biopharmaceuticals manufacturing and supply chains. It can improve the economy and public safety, especially during the COVID-19 pandemic. 



\section*{Acknowledgment}
We acknowledge the insightful discussion with Mike Baker from Willamette Valley Assured LLC, and the significant contribution of mobile App development from Fei Wang, Rui Li, Wen Li, Wei Su in San Jose State University. We also acknowledge the support of blockchain-enabled IoT platform validation from Dr. Jeffrey Steiner, Dr. Richard Roseberg, Dr. Gordon B. Jones, Chris Ringo, and Kristin Rifai from Oregon State University, Dr. Chad A. Kinney, Dr. Sang Hyuck Park from Colorado State University-Pueblo, Dr. Blake Osborn from Colorado State University, Dr. Ernst Cebert, Dr. Xianyan Kuang from Alabama A\&M University. 

\bibliographystyle{IEEEtran}
\bibliography{IEEEmybib}




%

\begin{IEEEbiography}[{\includegraphics[width=1in,height=1.25in,clip,keepaspectratio]{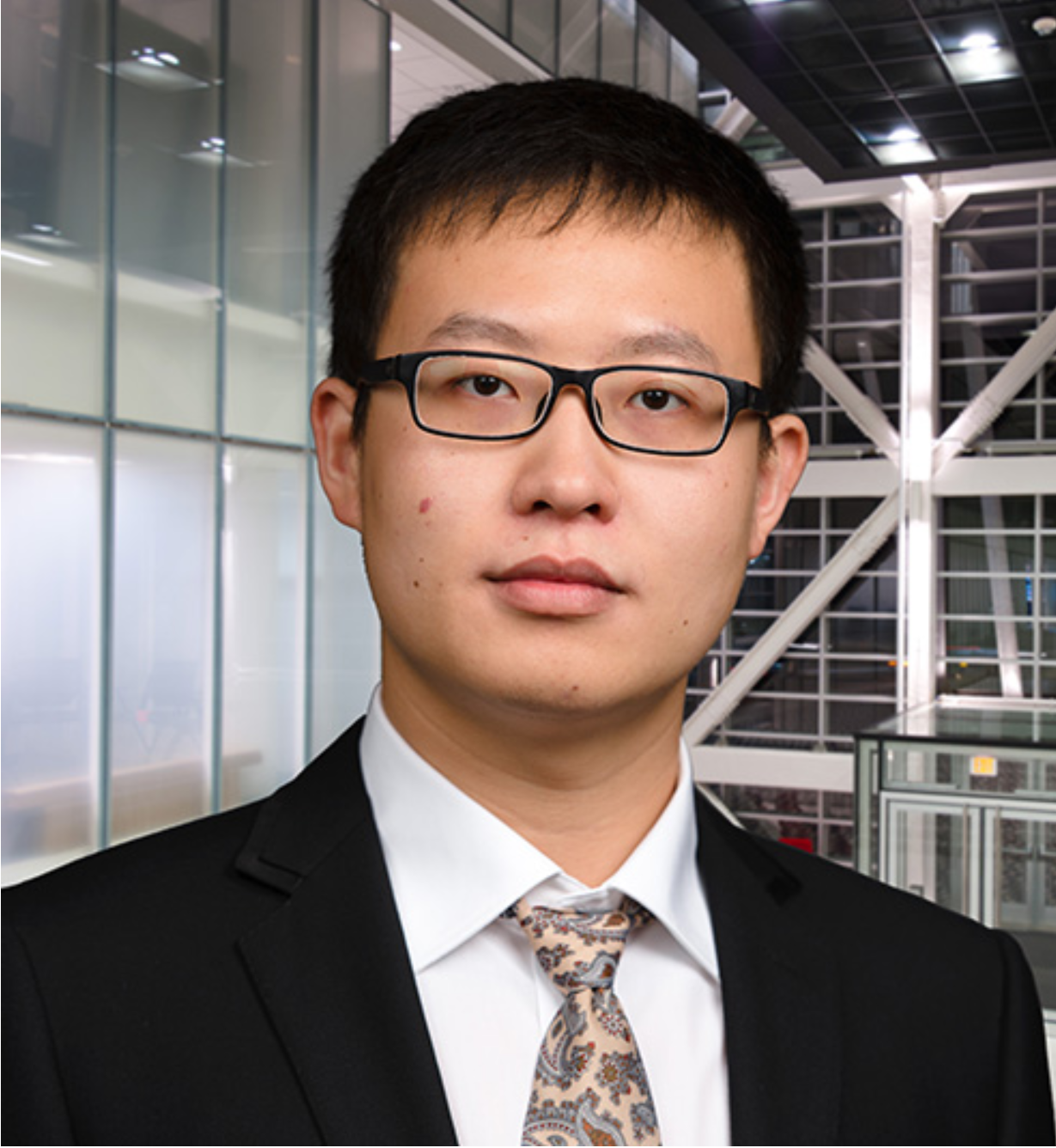}}]{Keqi Wang} is a Ph.D. student of the Department of Mechanical and Industrial Engineering (MIE) at Northeastern University (NEU).
He received his M.S. from Rutgers University in 2018 and B.S. from Shanghai University of Finance and Economics in 2014. Prior to joining NEU, he was a Trader at Coefficient Investment. His research interests are blockchain, stochastic computer simulation, data analytics and risk management.
\end{IEEEbiography}

\begin{IEEEbiography}[{\includegraphics[width=1in,height=1.25in,clip,keepaspectratio]{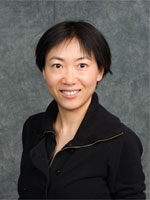}}]{Wei Xie} is an Assistant Professor in Northeastern University. She received her Ph.D. degree on Industrial Engineering and Management Sciences from Northwestern University 2014. Her research interests focus on interpretable Artificial Intelligence (AI), internet-of-things (IoT), computer simulation, data integrity and data analytics, design of experiments, model-based reinforcement learning, data-driven stochastic optimization, digital twin and blockchain development for end-to-end cyber-physical system learning and risk management with applications, including biopharmaceuticals manufacturing, industrial hemp production and supply chains. Dr. Xie received the 2015 Outstanding Publication Award from the INFORMS Simulation Society. She currently serves as Associate Editor for ACM Transactions on Modeling and Computer Simulation and Technical Activity Committee (TAC) for National Institute for Innovation in Manufacturing Biopharmaceuticals (NIIMBL). 
\end{IEEEbiography}

\begin{IEEEbiography}[{\includegraphics[width=1in,height=1.25in,clip,keepaspectratio]{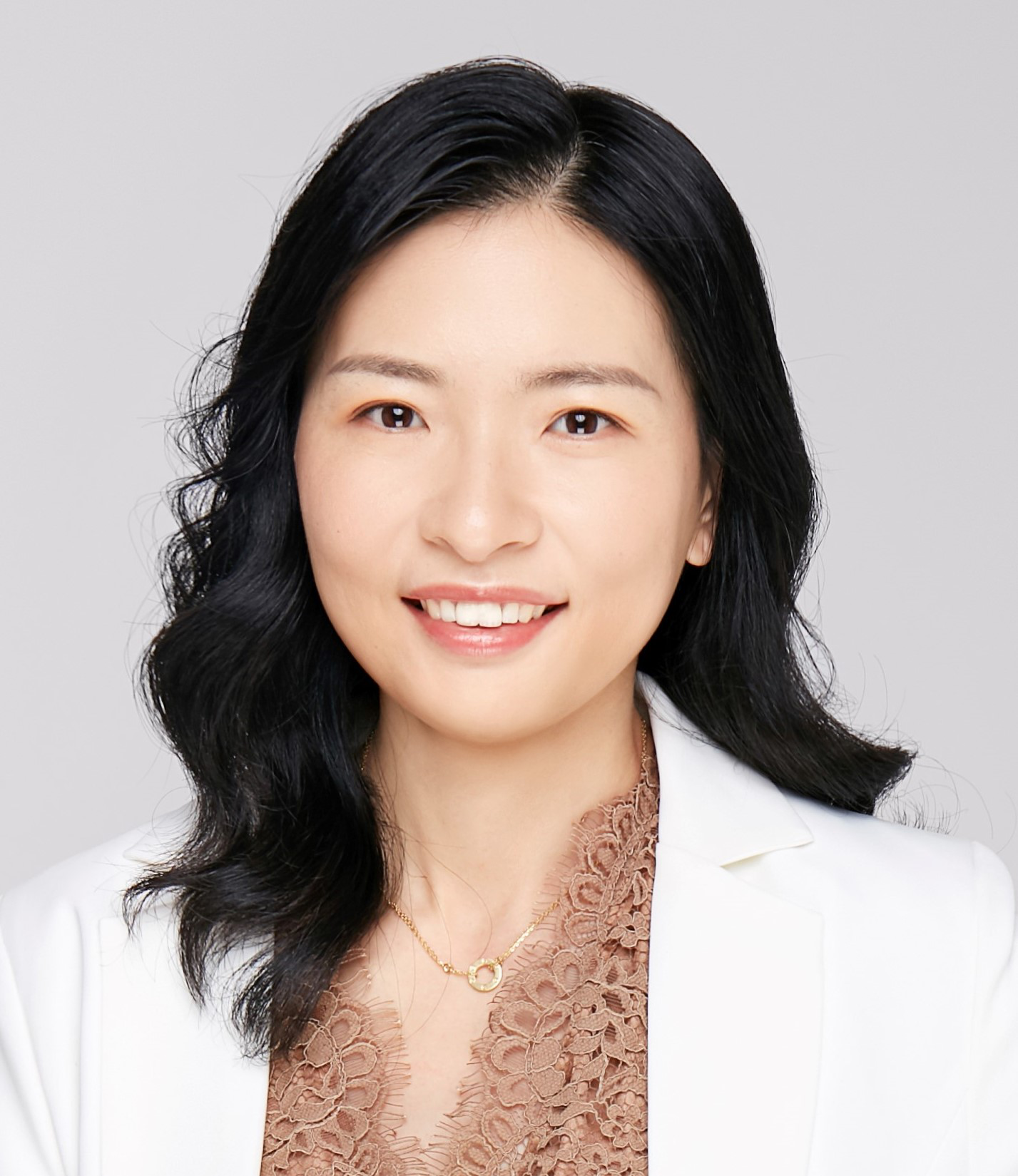}}]{Wencen Wu} is an Assistant Professor in the Computer Engineering Department at San Jose State University. Prior to joining SJSU, she was an Assistant Professor in the ECSE Department of Rensselaer Polytechnic Institute from 2013 - 2018. She received her Ph.D. and M.S. from the Georgia Institute of Technology in 2013 and 2010, and the dual-M.S. and B.S. from Shanghai Jiao Tong University. Her research interests include robotics, systems and control theory, and artificial intelligence applied to intelligent autonomous multi-robot systems and distributed parameter systems, and the applications of blockchain technology. 
\end{IEEEbiography}

\vspace{0 in}
\begin{IEEEbiography}[{\includegraphics[width=1in,height=1.25in,clip,keepaspectratio]{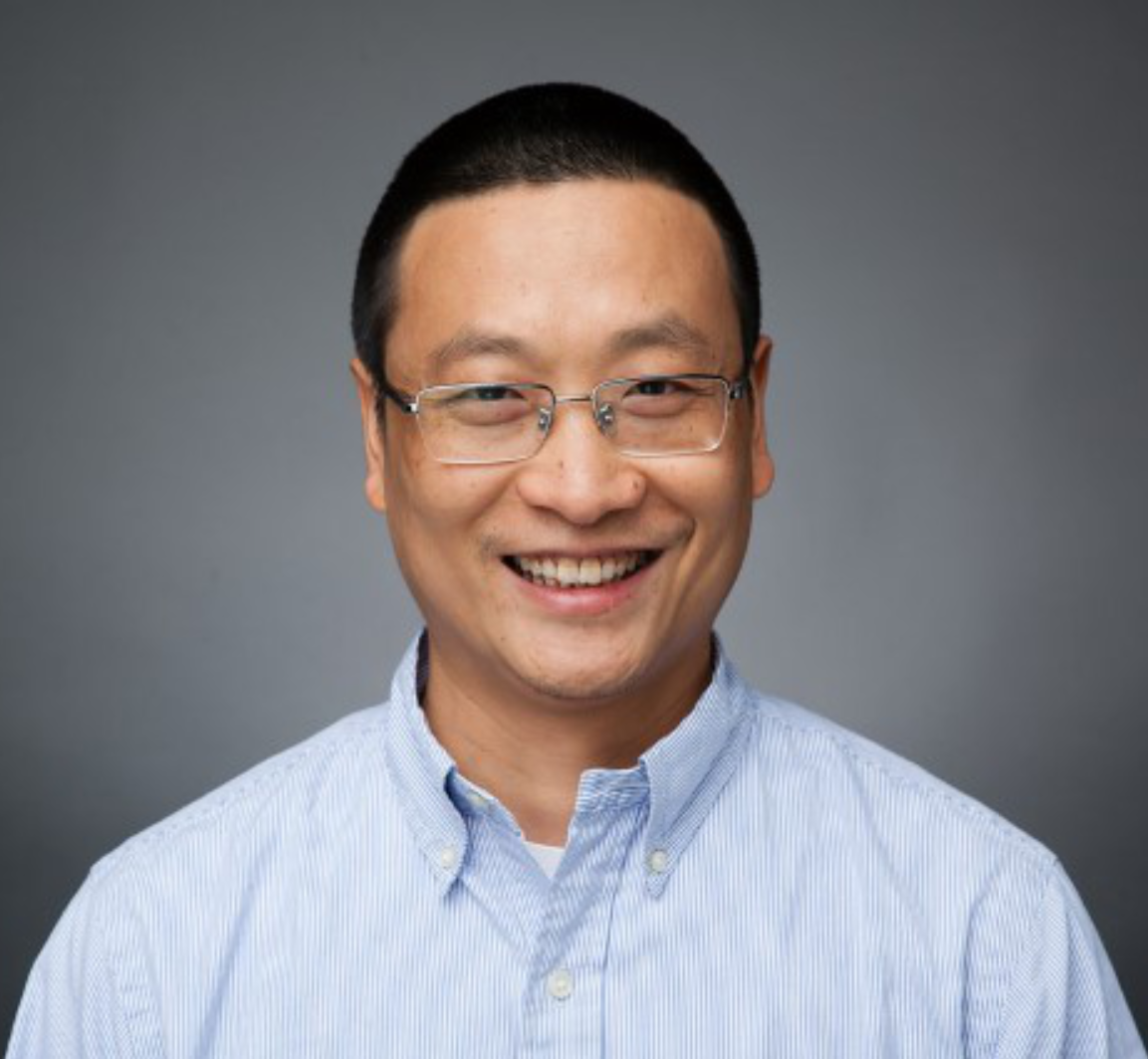}}]{Jinxiang Pei} is an Assistant Teaching Professor in the Department of Mechanical \& Industrial Engineering at Northeastern University. He received his M.S. and Ph.D. in Industrial Engineering and Management Sciences at Northwestern University. Prior to joining Northeastern, he was a Customer Demand Planning Manager at Nestle Waters NA and Demand Manager at Kraft responsible for improving organizational capability in statistical forecasting. His research interests are machine learning, supply chain management, game theory, and blockchain.
\end{IEEEbiography}

\vspace{0 in}
\begin{IEEEbiography}[{\includegraphics[width=1in,height=1.25in,clip,keepaspectratio]{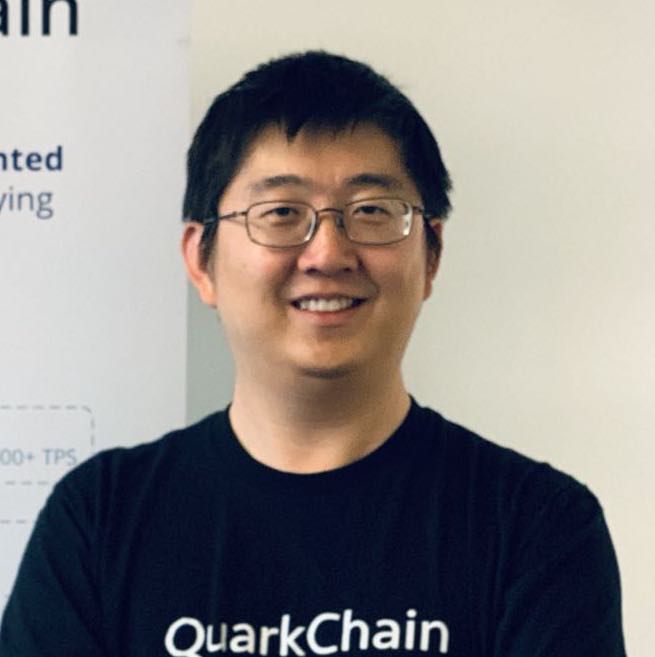}}]{Qi Zhou} is the Founder of QuarkChain Inc. Prior to founding QuarkChain in 2018, he worked as senior software engineer at Facebook and Google. He received his Ph. D. from the Georgia Institute of Technology in 2013, M.S. from Shanghai Jiao Tong University, and B.S. from Beijing University of Posts and Telecommunications. His research interests include blockchain, crypotocurrency, and large-scale distributed systems.
\end{IEEEbiography}

\end{document}